\documentclass[notitlepage,onecolumn,footinbib,floatfix,aps,10pt,pra,superscriptaddress]{revtex4-1}

\usepackage[usenames,dvipsnames]{xcolor} 
\usepackage{amssymb}
\usepackage{mathtools}
\usepackage{amsfonts}
\usepackage{bm}
\usepackage{braket}
\usepackage{bbm}
\usepackage{color}
\usepackage{latexsym}
\usepackage[english]{babel}
\usepackage{stmaryrd}
\usepackage{psfrag}
\usepackage{graphicx}
\usepackage{subfigure}
\usepackage{appendix}
\usepackage[utf8]{inputenc}
\usepackage[normalem]{ulem}
\usepackage{amsthm}

\definecolor{mygrey}{gray}{0.35}
\definecolor{myblue}{rgb}{0.2,0.2,0.8}
\definecolor{myzard}{cmyk}{0,0,0.05,0}
\definecolor{mywhite}{rgb}{1,1,1}
\definecolor{myred}{rgb}{0.9,0.1,0.}

\newcommand{\eref}[1]{(\ref{#1})}

\newcommand{\Fref}[1]{Fig.\ref{#1}}

\newcommand{\llrrq}[1]{\ensuremath{\left[ #1\right]}}

\DeclareMathOperator{\Tr}{Tr}
\renewcommand{\d}{\mathrm{d}}

\begin{document}

\title{Excitation dynamics in chain-mapped environments}

\author{Dario Tamascelli  }

\affiliation{%
 \quad Dipartimento di Fisica ``Aldo Pontremoli'', Universit\`a degli Studi di Milano, via Celoria 16, 20133 Milano, Italy; dario.tamascelli@unimi.it
}
\affiliation{
 \quad Institut f\"{u}r Theoretische Physik  and Center for Integrated Quantum Science and Technology (IQST), Albert-Einstein-Allee 11, Universit\"{a}t Ulm, 89069 Ulm, Germany
}

\begin{abstract}
The chain mapping of structured environments is a most powerful tool for the simulation of open quantum system dynamics. Once the environmental bosonic or fermionic degrees of freedom are unitarily rearranged into a one dimensional structure, the full power of Density Matrix Renormalization Group (DMRG) can be exploited.  Beside resulting in  efficient and numerically exact simulations of open quantum systems dynamics, chain mapping provides an unique perspective on the environment: the interaction between the system and the environment creates perturbations that travel along the one dimensional environment at a finite speed, thus providing a natural notion of light-, or causal-, cone. In this work we investigate the transport of excitations in a chain-mapped bosonic environment. In particular, we explore the relation between the environmental spectral density shape, parameters and temperature, and the dynamics of excitations along the corresponding linear chains of quantum harmonic oscillators. Our analysis unveils fundamental features of the environment evolution, such as localization, percolation and the onset of stationary currents.
\end{abstract}





\maketitle

\section{Introduction}
The thorough understanding of transport of energy, heat, particle, or mass in complex quantum systems is of utmost importance both from a fundamental and technological point of view . Such a relevance is witnessed by the enormous efforts invested by the scientific community over the last decades on the  theoretical and experimental investigation of the unique features of transport at the quantum regime. 

A variety of different topics can be put under the umbrella of quantum transport, such as efficient energy transfer and conversion in biological systems \cite{may04,plenio08,caruso09,chin13,tama19PRA}, transport in low dimensional quantum systems \cite{anderson58,topinka01,baer04, hartmann04,rossini09,osellame13}, quantum thermodynamics \cite{anufriev17,matti15}, and quantum information processing and transmission\cite{feyn82,childs03,tama13,benedetti19}.

Open quantum systems (OQS) formalism  \cite{breuer02, weiss12} has been widely employed for the description of quantum transport in the, often unavoidable, presence of additional and uncontrollable degrees of freedom interacting with the system under study. The tools provided by open quantum system theory led to the derivation of fundamental results allowing to understand and control, or at least mitigate, environmental effects. Such control, for example, is of utmost importance to preserve the quantum resources, as entanglement and coherence, that could enable the development of quantum devices possibly outperforming their classical counterparts. On the other side, the analysis of certain open quantum systems has unveiled the delicate interplay between coherence and sources of decoherence, as in the paradigmatic case of energy transport in disordered lattices \cite{plenio08,caruso09,tama13,tama17}.

The simulation of open quantum systems, on the other hand, represents a formidable task. Even when a microscopic description of the environment surrounding a quantum system is available, the derivation of the open quantum system  dynamics requires the solution of a number of differential equations that scales exponentially with the number of environmental
degrees of freedom. Analytic solutions are not available but for very few cases~\cite{luczka90,hu92,garraway97,fisher07,smirne10,diosi14,ferialdi16} and numerical
integration is not feasible, unless more or less severe approximations are used. Such
approximations, however, may fail to capture the effects of the interaction of open systems with environments that are either structured, or evolve on time-scales comparable to those characteristic of the open system. Electronic excitation or electron transport in a vibrational environment, ubiquitous in solid state environments and bio-molecular systems~\cite{huelga13,rivas14,breuer16,deVega17}, is just an example of this class of problems, which are of fundamental importance in a broad range of fields including the emergent quantum technology.

Over the last two decades, a variety of numerically exact approaches for the simulation of open quantum systems have been proposed \cite{deVega17}.  These methods allow for the description of features that were not accurately described by approximate methods, such as the Markov, Bloch-Redfield or perturbative expansion techniques \cite{breuer02}.  Among them we mention the hierarchical equations of motion (HEOM) \cite{kubo89,ishizaki09,tanimura06}, path integral methods \cite{feynman48,makri92,NalbachET2010}, Dissipation-Assisted Matrix Product Factorization \cite{somoza19}, and pseudo-modes related transformations \cite{garraway97,mascherpa20,lambert19}.

Time  Evolving  Density  operator  with  Orthogonal Polynomials (TEDOPA) \cite{prior10,chin10} algorithm is a method for the non perturbative simulation of OQS. TEDOPA  has  been employed to study a variety of open quantum systems \cite{prior10,prior13,chin13}.  TEDOPA belongs to the class of chain-mapping techniques  \cite{hughes09,prior10,chin10, martinazzo11,woods14,ferialdi15}, based on a unitary mapping of the environmental modes onto a chain of harmonic oscillators with nearest-neighbor interactions.  The main advantage of this mapping is the more local entanglement structure which allows for a straightforward application of density matrix renormalization group (DMRG) methods \cite{white92}. Moreover, the availability of bounds on the numerical errors introduced by the DMRG parametrization   allows to certify the accuracy of the results generated by TEDOPA  \cite{woods15}. 

As we will show starting from the next section, after the transformation of the environmental degrees of freedom into a linear chain of bosonic modes, the open system interacts only with the first site of the chain, where it dynamically creates (and destroys) excitations that subsequently propagate along the linear chain. A deeper understanding of excitation transport on bosonic chains obtained via the unitary chain mapping transformation of a bosonic environment can shed light on the mechanism that allows a linear structure to induce on the system the same dynamics of the original environmental configuration, where each oscillator was directly interacting with the system. The same linear structure, moreover, offers a unique point of view on the perturbations induced on the environment by the interaction with the system, since it naturally introduces a hierarchy of modes over which such perturbation propagate, or light-cone.

The paper is organized as follows. In Section \ref{sec:TEDOPA} we briefly introduce the TEDOPA chain mapping and fix our notation. In Section \ref{sec:chaindyn} we discuss the dynamical features of transport on TEDOPA chain associated, respectively, to Lorentzian and Ohmic spectral densities in the single excitation subspace. In Section \ref{sec:fulldyn} we extend the analysis by including the interaction with the open system. Section \ref{sec:Conclusion} is devoted to conclusion and outlook.

\section{TEDOPA} \label{sec:TEDOPA}
Here and in what follows we consider a system interacting with a bosonic environment. The complete  Hamiltonian reads ($\hbar = 1$):
\begin{align}
    H & =\label{eq:totHam} 
H_S + H_E+H_I   \\   
H_E &=  \int \d \omega   \omega a_\omega^\dagger a_\omega \nonumber \\
    H_I &=  A_S \int \d \omega  h(\omega) O_\omega, \nonumber
\end{align}
where $H_S$ is the (arbitrary) free system Hamiltonian, $H_E$ describes the free evolution of the bosonic environmental degrees of freedom, and $H_I$  is the bilinear system -environment interaction Hamiltonian \cite{leggett87}, and $A_S$, $O_\omega$ are self-adjoint operators. This last assumption is necessary for the Thermalized-TEDOPA (T-TEDOPA) mapping~\cite{tama19}, that we will introduce later in this section.  We assume that $h(\omega)$ has finite support $[\omega_\text{min},\omega_\text{max}]$, with $\omega_\text{min} <\omega_\text{max}$,
and define the spectral density (SD), namely the positive valued function $J:[\omega_\text{min},\omega_\text{max}] \to \mathbb{R}^+$, as
\begin{equation}
    J(\omega) =  h^2(\omega).
    \label{eq:sd}
\end{equation}

As shown in
\cite{chin10,prior10,woods14} the Hamiltonian \eref{eq:totHam} can be unitarily mapped into an equivalent one through
by defining a countably infinite set of new operators
\begin{align}
b_n^\dagger & = \int_{\omega_\text{min}}^{\omega_\text{max}} \d\omega U_n(\omega) a_\omega^\dagger \\
b_n & = \int_{\omega_\text{min}}^{\omega_\text{max}} \d\omega U_n(\omega) a_\omega
    \label{eq:modesTrans}
\end{align}
where 
\begin{equation}
    U_n(\omega) = h(\omega) p_n(\omega).
    \label{eq:unit}
\end{equation}
The operators $b_n$ and $b_n^\dagger$ satisfy the bosonic commutation relations
$[b_n,b_m^\dagger]=\delta_{nm}$; moreover, the polynomials $p_n(\omega)$ are orthogonal with respect to the measure $d \mu = J(\omega)
\d\omega$ and satisfy three-term recursion relations \cite{chin10,woods14}. Thanks to these properties, the Hamiltonian \eref{eq:totHam} is mapped into the one dimensional Hamiltonian
\begin{align}
    H^C &= H_S + \kappa_0 A_s (b_1+b_1^\dagger) + \\
    & \sum_{n=1}^{+\infty} \omega_n b_n^\dagger b_n +  \kappa_n(
    b_{n+1}^\dagger b_n + b_n^\dagger b_{n+1}) \nonumber \\
    & = H_S + H_I^C + H_E^C.
    \label{eq:chainHam}
\end{align}
where, for the sake of definiteness, we have specialized the operator $O_\omega$  in \eref{eq:totHam} to $X_\omega = a_\omega + a_\omega^\dagger$.
After the mapping, the system interacts with the first site of a linear (infinite) chain of bosonic modes;  the system-chain interaction strength is given by \cite{chin10,woods14}
\begin{equation} 
    \kappa_0^2 = \int_{\omega_\text{min}}^{\omega_\text{max}} \d \omega J(\omega), \label{eq:overall}
\end{equation}
whereas the frequency of the first TEDOPA chain oscillator is 
\begin{equation}
    \omega_1 = \int_{\omega_\text{min}}^{\omega_\text{max}} \d \omega  \omega\frac{ J(\omega)}{\kappa_0^2},
\end{equation}
namely the first moment of the normalized measure $J(\omega)/\kappa_0^2 \d \omega$ on $[\omega_\text{min},\omega_\text{max}]$.
The remaining coefficients $\omega_n$ and $\kappa_n$ are defined by the above mentioned three-terms recursion relations; while in certain cases  it is possible to analytically determine their value \cite{chin10}, a numerically stable procedure is in general used \cite{gautschi94,gautschi04}. 

For the following analysis, it is important to stress that the chain Hamiltonian $H_E^C$ is made up of exchange terms $b_{n-1} b_n^\dagger + H.c.$ and therefore conserves the
``number'' operator, i.e. $[N,H_E^C]=0$ where
\begin{equation}
    N = \bigotimes_{n=1}^\infty b_n^\dagger b_n.
    \label{eq:numberOp}
\end{equation}
Excitations can therefore be added or subtracted from the chain because of the interaction with the system.

The initial joint system-environment state is assumed factorized $\rho_{SE}(0) = \rho_S(0) \otimes \rho_{E,\beta}(0)$, with  $\rho_{E,\beta}(0)$ a thermal state at inverse temperature $\beta = 1/k_B T$, namely 
\begin{align} \label{eq:thermalEnv}
\rho_{E,\beta}(0) = \bigotimes_\omega  \exp(-\beta\omega a^\dagger_\omega a_\omega) /\mathcal{Z}_\omega,
\end{align}
with $\mathcal{Z}_\omega = \Tr[\exp(-\beta\omega a^\dagger_\omega a_\omega)]$ the partition function. The initial state after the chain mapping is a factorized state $\rho_{SE}^C(0) = \rho_S(0) \otimes \rho_{E,\beta}^C(0)$ as well with
\begin{align} \label{eq:thermalEnvChain}
    \rho_{E,\beta}^C = \exp(-\beta H_E^C)/\mathcal{Z}_E^C.
\end{align}
If the environment is initially at zero temperature, its initial state is the vacuum state, and the initial state of the chain is also a factorized vacuum state $\ket{0}_E^C$ (i.e. $b_k \ket{0}_E^C = 0, k=1,2,\ldots$): the chain contains therefore no excitations. This case provides us with the simplest setting where to analyze the transport properties of the chain corresponding to some representative spectral densities.
As recently shown in \cite{tama19}, however, by the spectral density transformation
\begin{align} 
    J_\beta(\omega) = \frac{J'(\omega) }{2}\left [  1+ \coth\left (\frac{\beta \omega}{2} \right )\right ] \label{eq:thermalizedSD}
\end{align}
with $J'(\omega) = \text{sign}(\omega) J(| \omega|)$, it is always possible to replace the thermal state of the original environment with the vacuum state of an extended environment, comprising negative frequencies. As the spectral density \eref{eq:thermalizedSD} is now temperature dependent, the TEDOPA chain coefficients $\omega_{n,\beta}, \ \kappa_{n,\beta}$ will be temperature dependent as well. In the following we will drop the $\beta$ dependence wherever clear from the context. From now on will therefore always consider the factorized vacuum state as the initial chain state without loss of generality.

In our analysis we will consider, in particular, the Lorentzian spectral density
\begin{align}
    J_L(\omega) =  \frac{\lambda^2}{\pi}  \frac{4 \gamma \Omega  
        \omega}{\llrrq{\gamma^2 +(\omega + \Omega)^2}\llrrq{ \gamma^2
        +(\omega -\Omega)^2}}, \label{eq:asymmLorentz}
\end{align}
and Ohmic spectral densities 
\begin{align}
    J_O^s(\omega) =  \frac{\lambda^2}{\pi} \frac{\omega^s}{s! \omega_c^{s-1}} e^{-\frac{\omega}{\omega_c}}, \label{eq:ohmicSD}
\end{align}
defining a very important class of environments entering in the study of many systems, such as microscopic models leading to a Lindblad master equation for an harmonic oscillator in a weakly coupled high temperature environment, or a particle undergoing quantum Brownian motion \cite{breuer02,leggett83,ford88}. From now on frequencies will be in cm$^{-1}$ and temperatures in Kelvin. We remark that, because of the relation \eref{eq:overall}, if two spectral densities differ only for the overall coupling constant $\lambda$,  their mappings (i.e. all of the chain coefficients $\omega_n, \kappa_n$) will be identical, with the exception of $\kappa_0$, namely the coupling strength between the system ant the first TEDOPA chain mode. We also observe that the chain coefficients $\omega_n, k_n, \  n \geq 1$ are independent of the specific system-environment interaction term, i.e. they are independent of the choice of $A_S, O_\omega$ of equation \eref{eq:totHam}. As customary for chain mappings, in what follows, we will moreover impose a hard cutoff $\omega_\text{hc}$ to the considered spectral densities, thus limiting their support to the interval $[0,\omega_\text{hc}]$ for $T=0$ and to the interval $[-\omega_\text{hc},\omega_\text{hc}]$ for $T>0$. The value of $\omega_\text{hc}$ is suitably chosen as to keep the neglected relative reorganization energy 
\begin{equation}
\frac{\int_{\omega_\text{hc}}^\infty \d \omega J(\omega)/\omega}{\int_{0}^{\infty} \d \omega J(\omega)/\omega} \nonumber
\end{equation}
in the order of $10^{-4}$ for all the considered instances.

If the considered spectral density belongs to the Szeg\"o class, the asymptotic relations
\begin{align} \label{eq:asymcoeff}
    \omega_\infty &= \lim_{n \to \infty} \omega_n = \frac{\omega_\text{max}+\omega_\text{min}}{2} \\
    \kappa_\infty &= \lim_{n \to \infty} \kappa_n = \frac{\omega_\text{max}-\omega_\text{min}}{4} \nonumber
\end{align}
hold (see Theorem 47 of \citet{woods14}). Clearly enough, in our setting $\omega_\text{max}$ (and, at finite temperature, $\omega_\text{min}$) depends on the imposed hard-cutoff $\omega_\text{hc}$ so that both $\omega_\infty$ and $\kappa_\infty$ are functions of $\omega_\text{hc}$. For any suitably fixed $\omega_\text{hc}$, however, the relations \eref{eq:asymcoeff} allow for the simple heuristic estimation $L = 2 \kappa_\infty t_\text{max}$ of the maximal distance travelled within the time $t_\text{max}$ by an excitation initially located at the first TEDOPA chain site. For fixed time $t_\text{max}$, therefore, the effective environment is made up of $L$ oscillators within the ``light-cone''. Interestingly enough, the width of such light cone depends only on the ``artificially'' imposed hard cutoff and, as long as the choice $\omega_\text{hc}$ is sensibly chosen, different choices of the hard-cutoffs do not impact on the reduced dynamics of the system. On the other side, different spectral densities with the same support will have the same asymptotic coefficients, and the differences in the reduced dynamics of the system will be due to a (typically quite small) finite number of modes, as we will see in the following sections.

\section{Chain Dynamics} \label{sec:chaindyn}
We start by analyzing the dynamics of a single excitation moving along the chain-mapped environment  produced by the (T-)TEDOPA mapping. To this end, we can disregard the system and the interaction term $H_I$, or equivalently set $\kappa_0=0$, and restrict our attention to the single excitation sector of the  TEDOPA-chain Hilbert space. The set $\left \{\ket{k},\ k=1,2,...\right \}$, where $\ket{k}$ indicates the Fock state $\ket{n_1 = 0,\ldots,n_{k-1}=0,n_k=1,n_{k+1}=0,\ldots} $ with the single excitation located at the $k$-th chain site, is a basis for the considered single excitation subspace. In what follows we will assume that the excitation is initially located at site $1$, namely the initial state is $\ket{1}$.


\subsection{Lorentzian spectrum.} \label{sec:lorentzSD}

The Lorentzian spectral density \eref{eq:asymmLorentz} provides a paradigmatic example. For $\gamma/\Omega \ll 1$, such spectrum well approximates that of an environment made up of a single harmonic oscillator with frequency $\Omega$ and dissipating into the vacuum at rate $\gamma$ \cite{breuer02,lemmer17}. In all the following examples a hard cutoff frequency $\omega_\text{hc} = 10\Omega$ has been enforced.
%
\begin{figure} 
\subfigure[]{\includegraphics[width=0.3\textwidth]{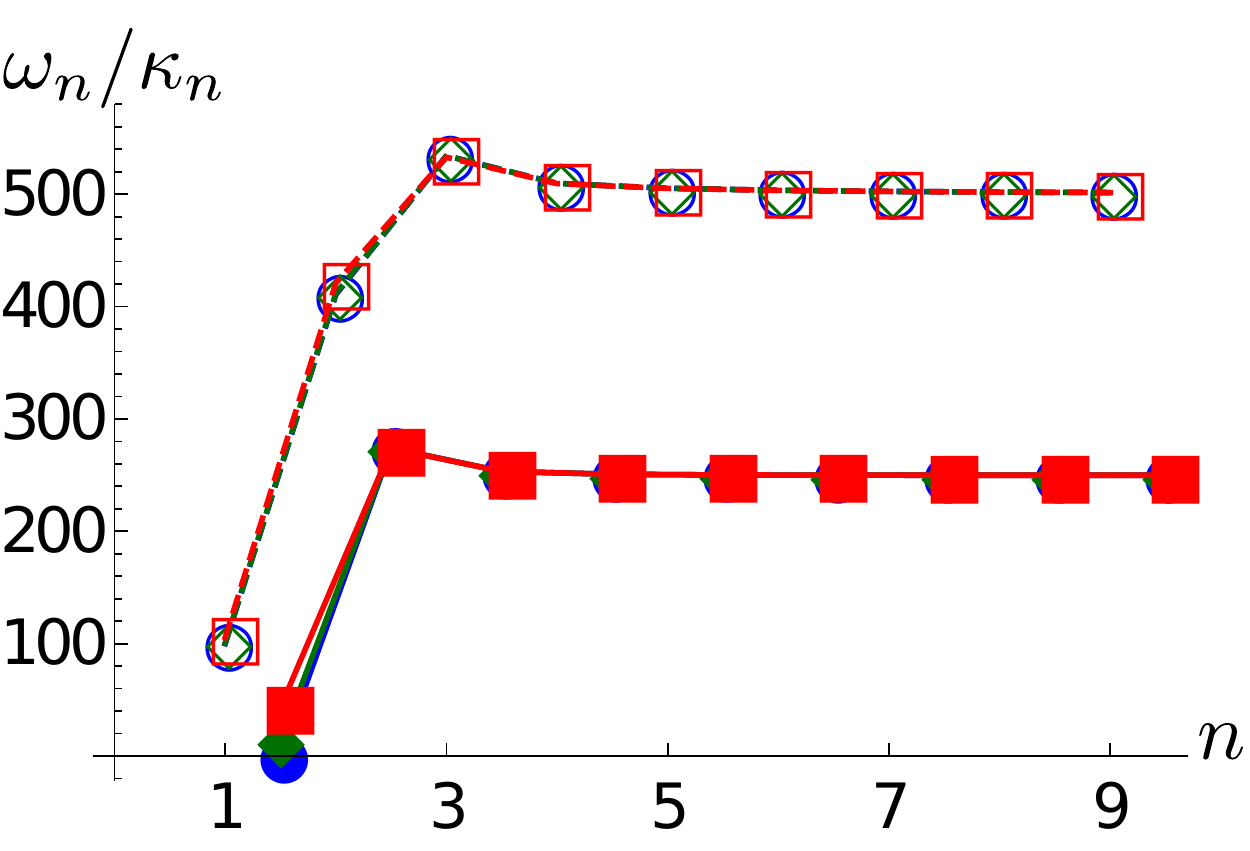}}
\subfigure[]{\includegraphics[width=0.3 \textwidth]{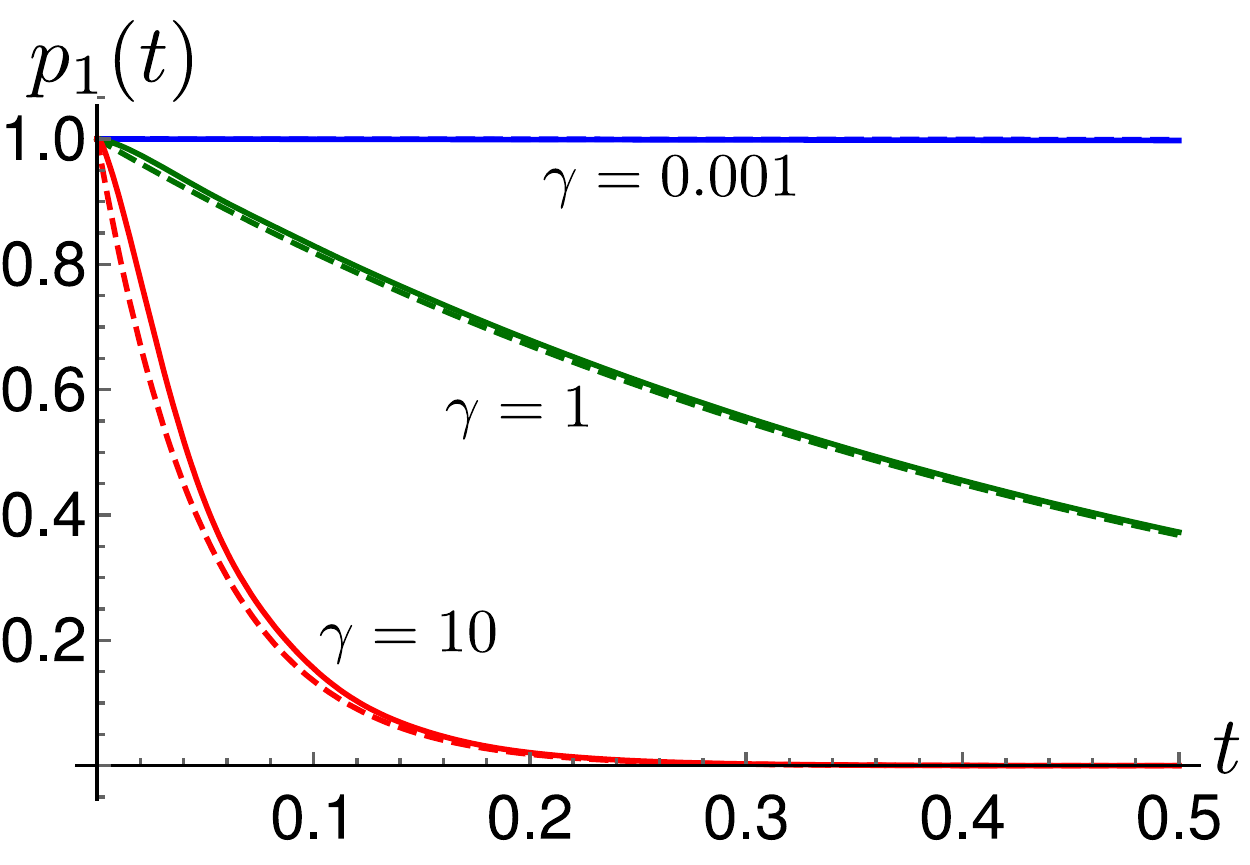}}
\subfigure[]{\includegraphics[width=0.3 \textwidth]{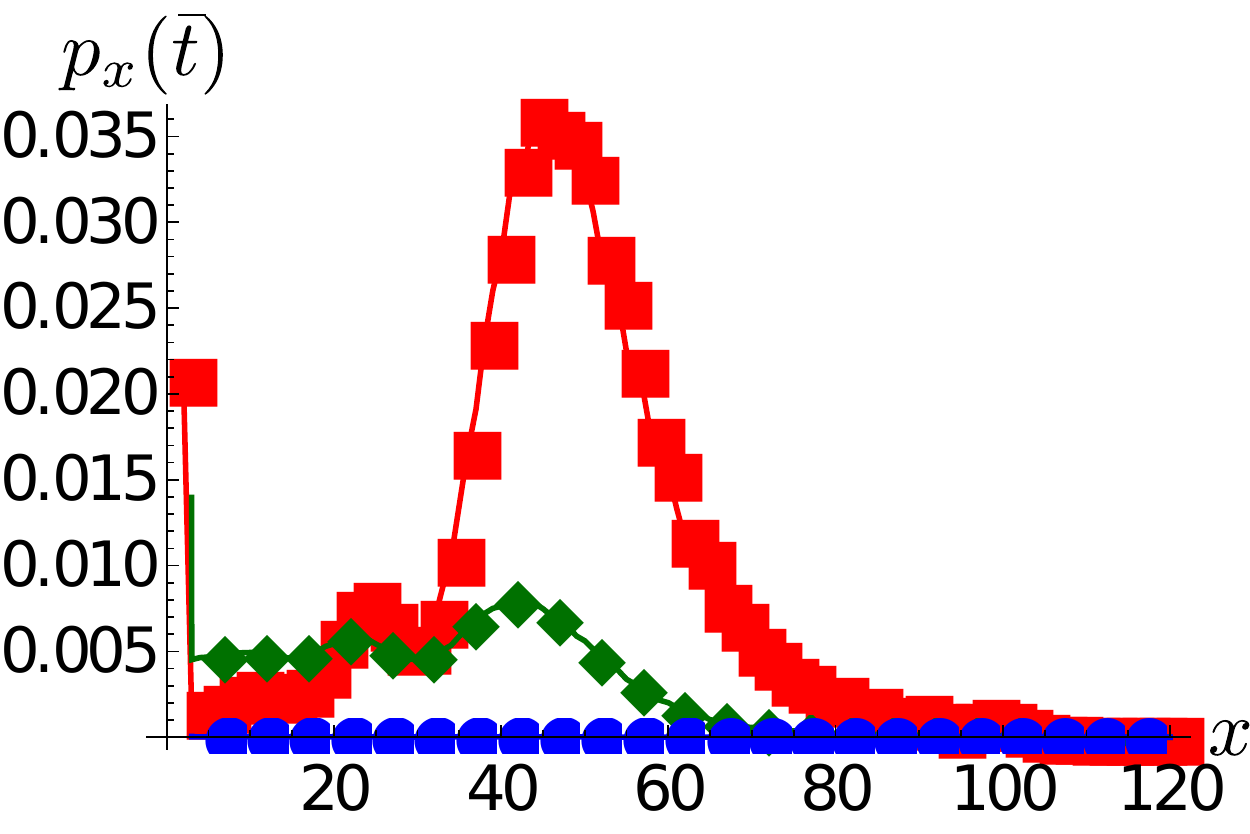}}
\caption{\label{fig:LorentzT0} Lorentzian SD; in all frames $\Omega_S=10$, $T=0$; blue, green and red lines/marker refer respectively to $\gamma=0.001$, $\gamma=1$ and $\gamma=10$. (a) The chain parameters $\omega_n$ (empty markers) and $\kappa_n$ (filled markers) for $\gamma=0.001$ (blue circles), $\gamma = 1$ (green diamonds) and $\gamma = 10$ (red squares); the couplings are shifted by $0.5$ to the right to lie between $n$ and $n+1$. (b) The population $p_1(t)$ of the first site as a function of time; the decay rates $\exp(-2 \gamma)$ are shown as dashed lines as a guide to the eye. (c) The population of $p_x(\bar{t})$ at $\bar{t}=0.2$ for $x=1,2,\ldots,120$.}
\end{figure}
Frame (a)  of \Fref{fig:LorentzT0} shows the frequencies $\omega_n$ and couplings coefficients  $\kappa_n$  at $T=0$ for $\Omega=100$ and $\gamma =0.001,1,10$ (see \eref{eq:asymmLorentz}). We first observe that, for all values of $\gamma$, the first and the second TEDOPA chain modes are equally far detuned. The main difference between the three selected cases lies  in the coupling strength $\kappa_1$ between the same two modes, which is directly proportional to $\gamma$. The effect on the system dynamics is remarkable. As shown in \Fref{fig:LorentzT0}(b) the population of the first TEDOPA chain is well approximated by $p_1(t) = \exp(-2 \gamma t)$, namely the decay rate of an harmonic oscillator damped into the vacuum at a rate $\gamma$.  As frame (c) of the same figure shows, the portion of excitation that propagates beyond the first site propagates on the TEDOPA chain at a speed which is independent of $\gamma$: the chain coefficients are essentially equal to each other in the three cases for $n \geq 3$, and their value is determined by the hard cutoff frequency $\omega_\text{hc}$ through \eref{eq:asymcoeff}.

We turn now our attention to the finite temperature case.

\begin{figure} 
\centering
\subfigure[]{ \includegraphics[width=0.3 \textwidth]{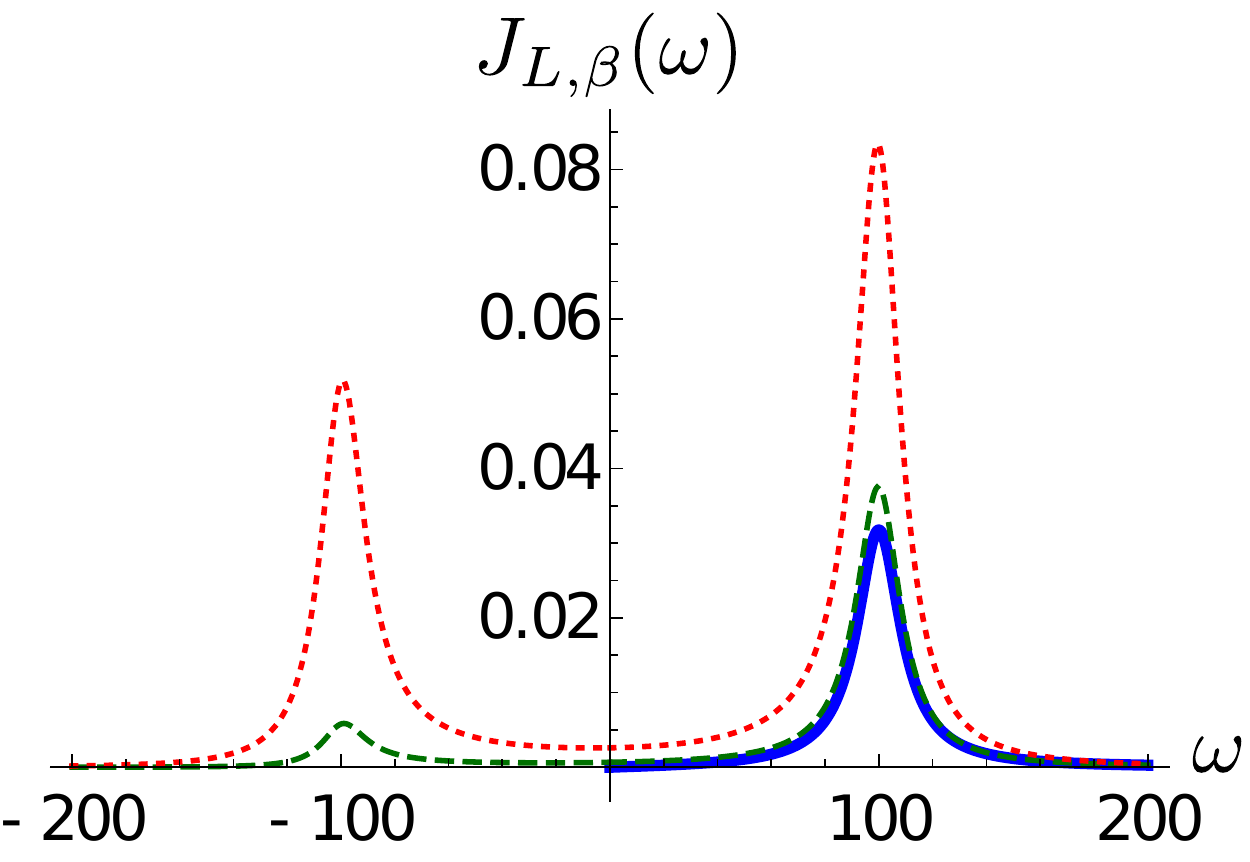}}
\subfigure[]{\includegraphics[width=0.3 \textwidth]{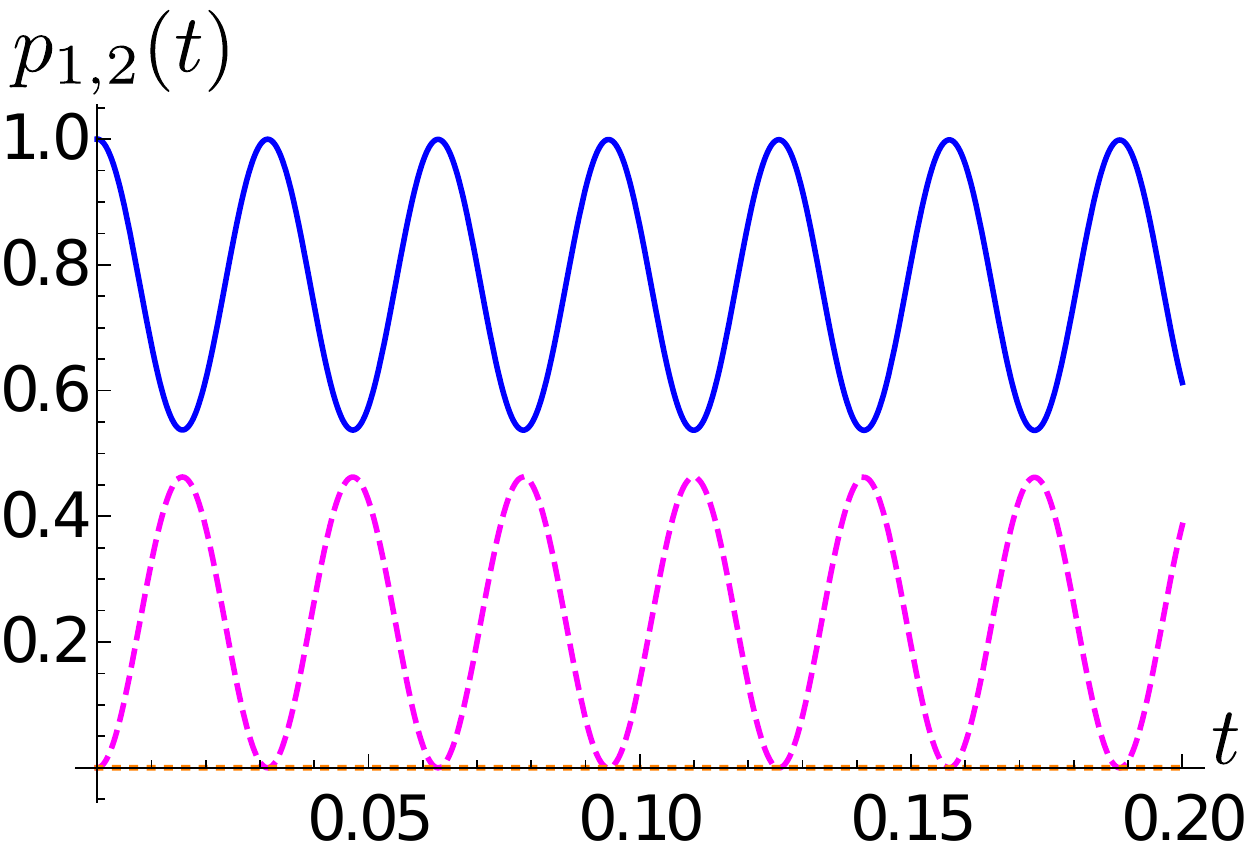}}
\subfigure[]{\includegraphics[width=0.3 \textwidth]{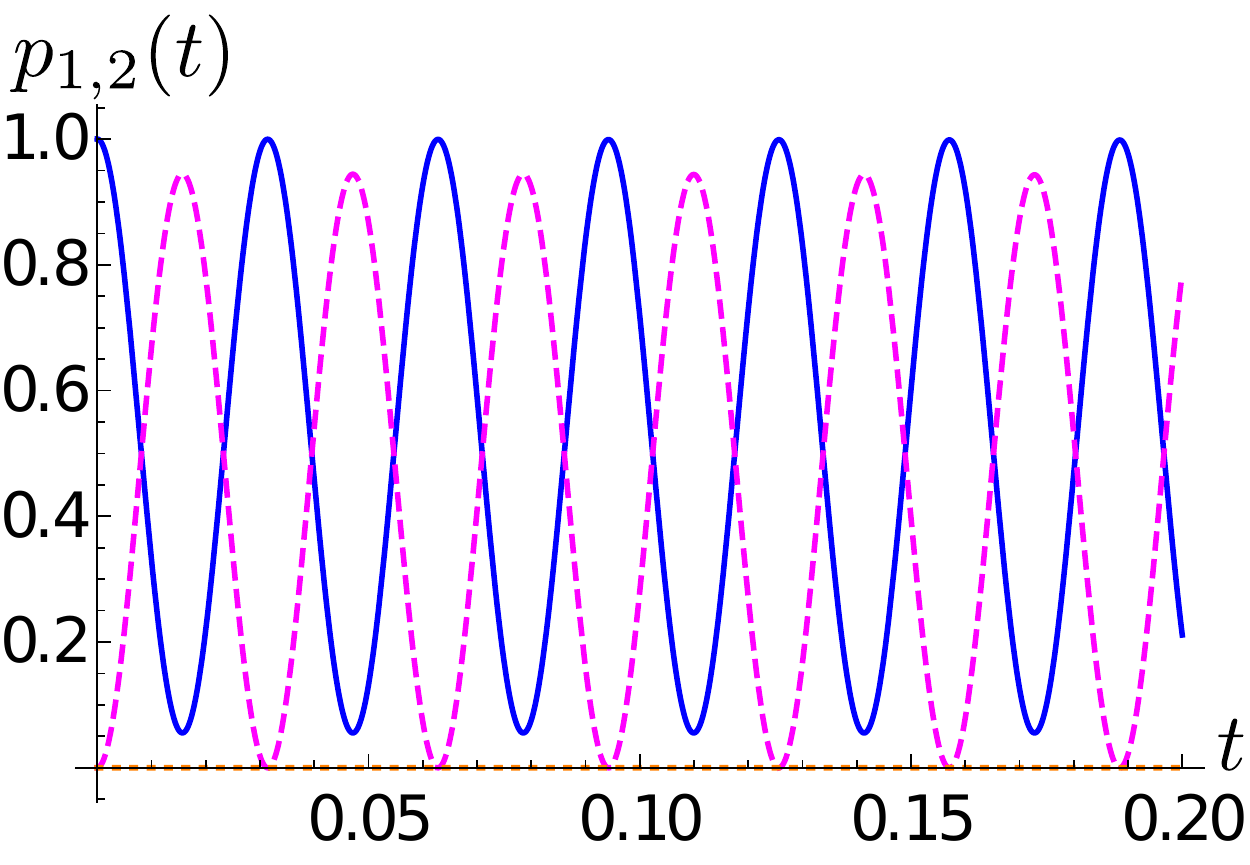}}\\
\subfigure[]{ \includegraphics[width=0.3 \textwidth]{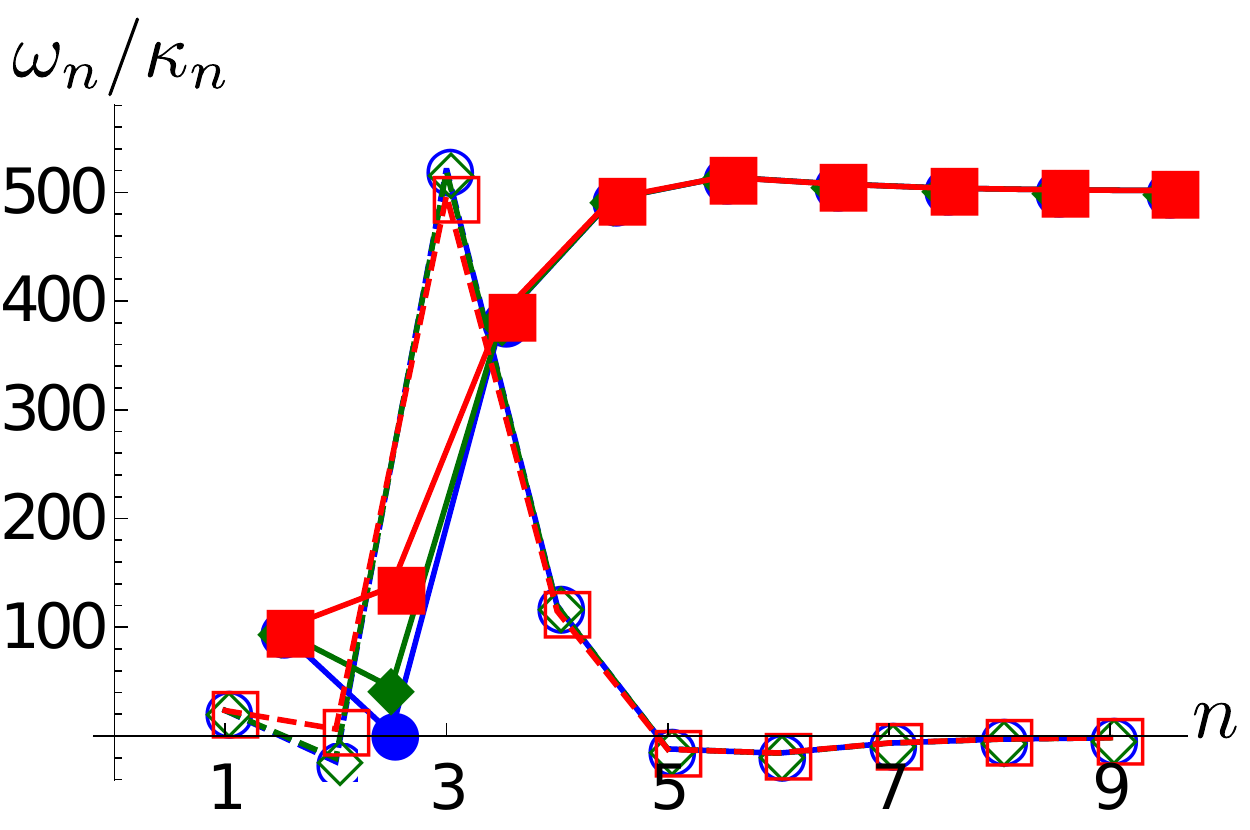}}
\subfigure[]{\includegraphics[width=0.3 \textwidth]{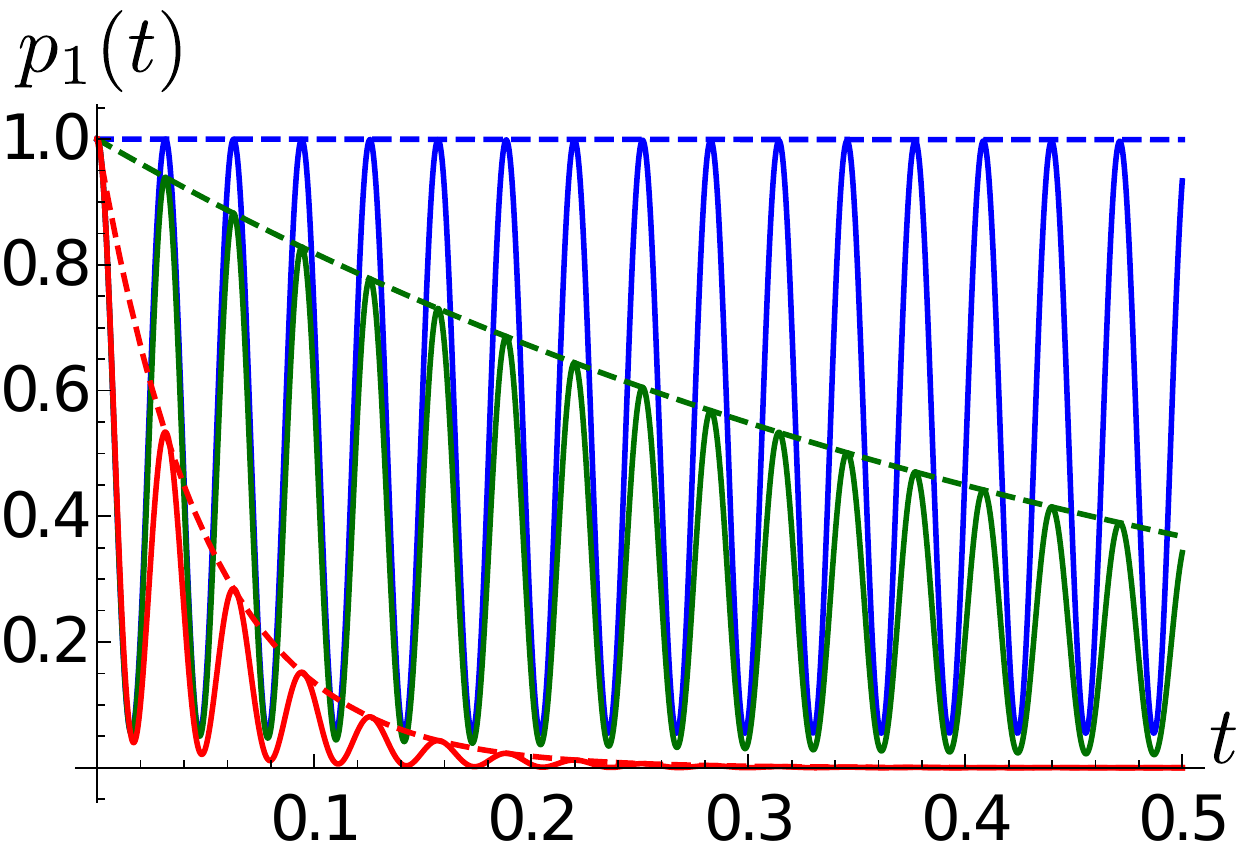}}
\subfigure[]{\includegraphics[width=0.3 \textwidth]{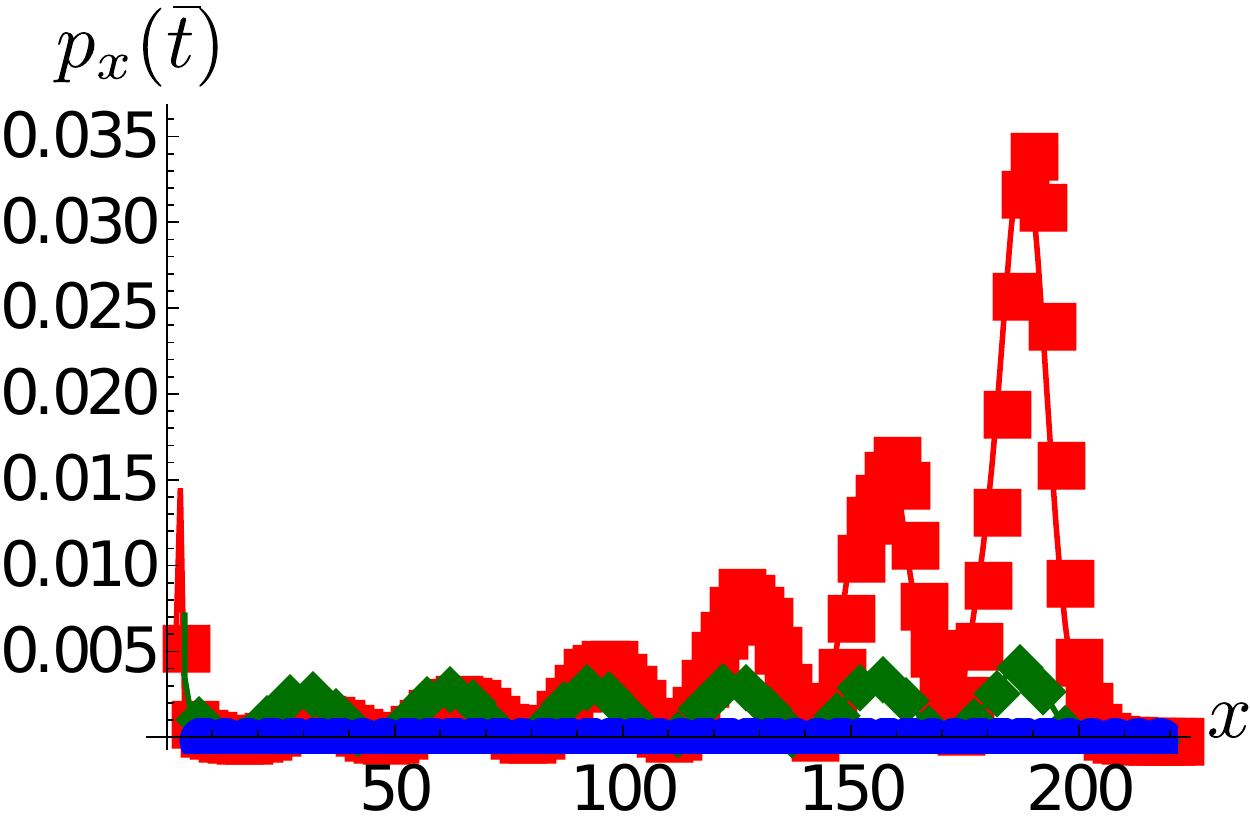}}
\caption{\label{fig:LorentzFT} (a) The thermalized Lorentzian SD $J_{L,\beta}(\omega)$ (see Eq. \ref{eq:thermalizedSD} and \ref{eq:asymmLorentz}) for $\Omega=100, \gamma = 10$ at $T=0$ (blue solid line), $T=77$ (green dashed line) and $T=300$ (red dotted line). In all the remaining frames $\Omega=10, \gamma = 0.001$. (b) $T=77$; the first (dashed blue line) and the second  (magenta dashed line) TEDOPA chain site populations $p_{1,2}(t)$ as a function of time. (c) Same quantities and line styles as frame (b) at $T=300$ (d)-(f): same quantities and styles as frames (a)-(c) of \Fref{fig:LorentzT0} for $T=300$.}
\end{figure}

As exemplified in \Fref{fig:LorentzFT}(a), after the thermalization procedure \cite{tama19} the thermalized  spectral density \eref{eq:thermalizedSD} presents two peaks at $\pm \Omega$. The system will be thus effectively coupled to two damped modes, with temperature dependent coupling strength proportional to $1+n_\beta(\Omega)$ resp. $n_\beta(\Omega)$.

It is thus not surprising that the chain dynamics for the case $\gamma=0.001$ is essentially confined to the first two chain modes, as frames (b) and (c) of \Fref{fig:LorentzFT} show. Indeed, the same plots suggest a clear relation between the temperature and the relative occupation of the modes: as $T$ increase, the difference between the maxima of the populations of the first and the second TEDOPA chain sites decreases, and is expected to vanish as $T \to \infty$, i.e. when the thermalized spectral density becomes symmetric with respect to the origin. 

It is interesting to see that a mechanism very similar to the one discussed for the zero temperature case is at play also at finite temperature. Frame (d) of \Fref{fig:LorentzFT} shows the chain coefficients for $\gamma=0.001,1$ and $10$ at $T=300$. This time the detuning between the first and the second TEDOPA chain sites is relatively small and the coupling between the two sites is independent of $\gamma$. This time it is the detuning between the second and the third chain site that is considerable, and the coupling $\kappa_2$ is monotone with $\gamma$. As shown in frame (e) of the same figure, the result is that the population of the first TEDOPA chain site presents damped beatings: the excitation moves forth and back between the first two chain sites, and percolates toward the right part of the chain at a rate $\exp(-2 \gamma t)$. In the zero temperature case, the ``escaped'' population travels toward the right part of the chain at a speed which is independent of $\gamma$, and keeps trace of such beatings, as shown in \Fref{fig:LorentzFT}(f), but this time the propagation speed is twice that of the zero temperature case because of the enlarged support $[-\omega_\text{hc},\omega_\text{hc}]$ (see \eref{eq:asymcoeff}).

\begin{figure} 
\centering
\subfigure[]{ \includegraphics[width=0.3 \textwidth]{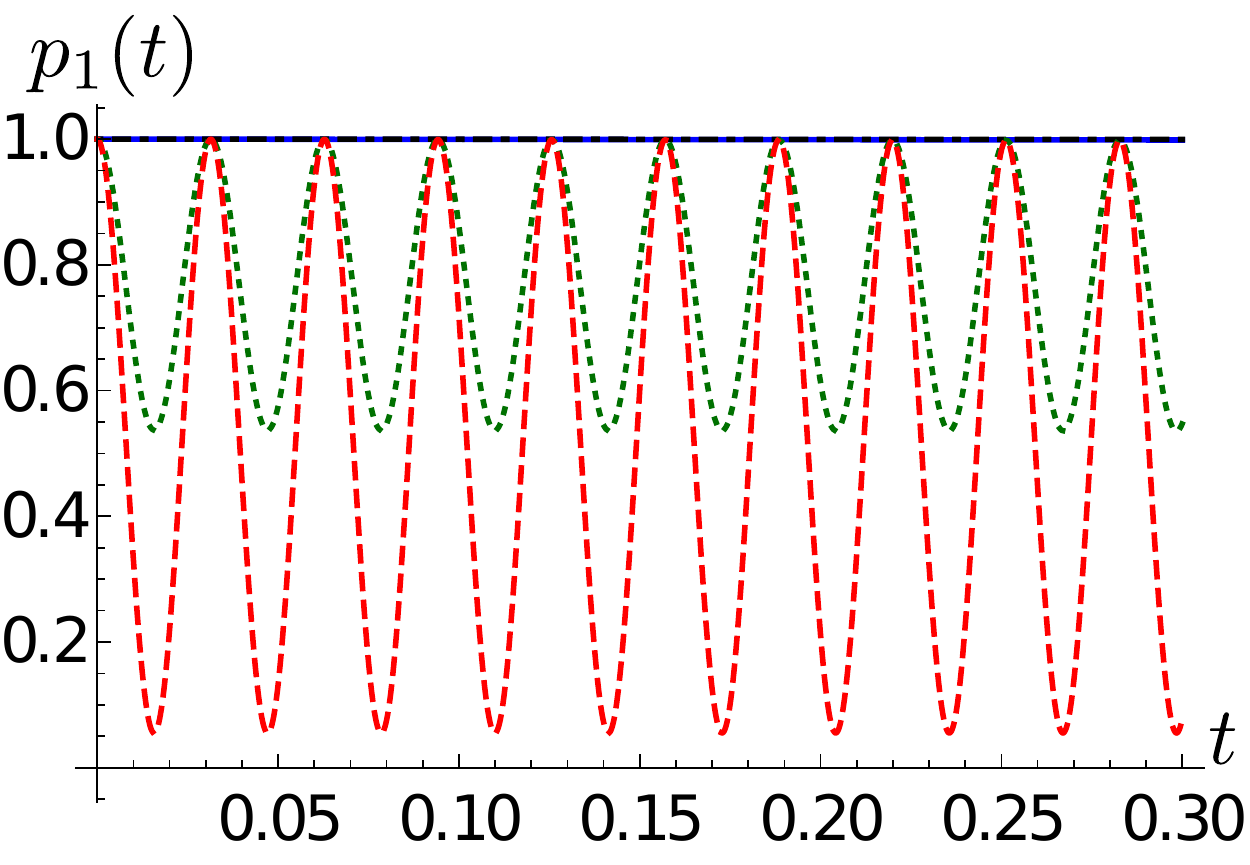}}
\subfigure[]{\includegraphics[width=0.3 \textwidth]{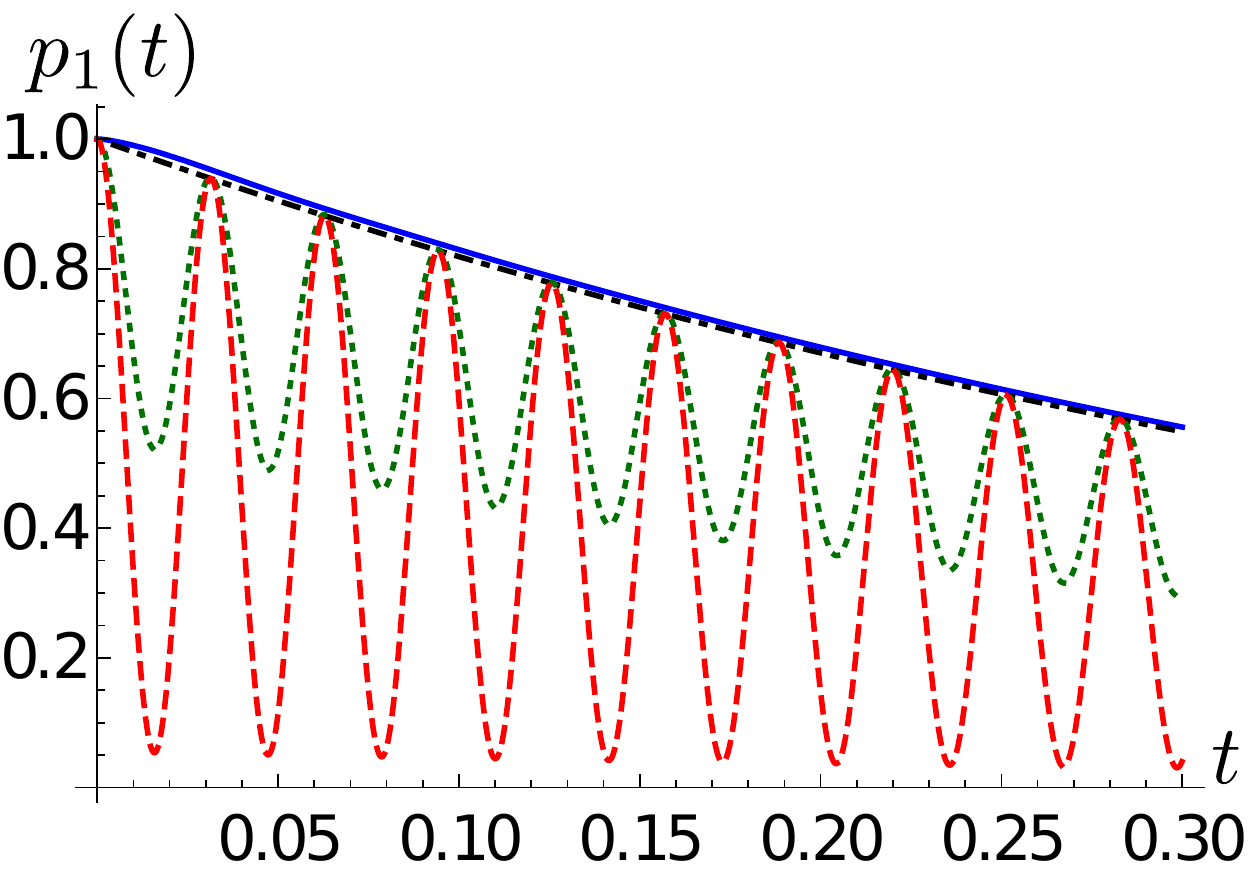}}
\subfigure[]{\includegraphics[width=0.3 \textwidth]{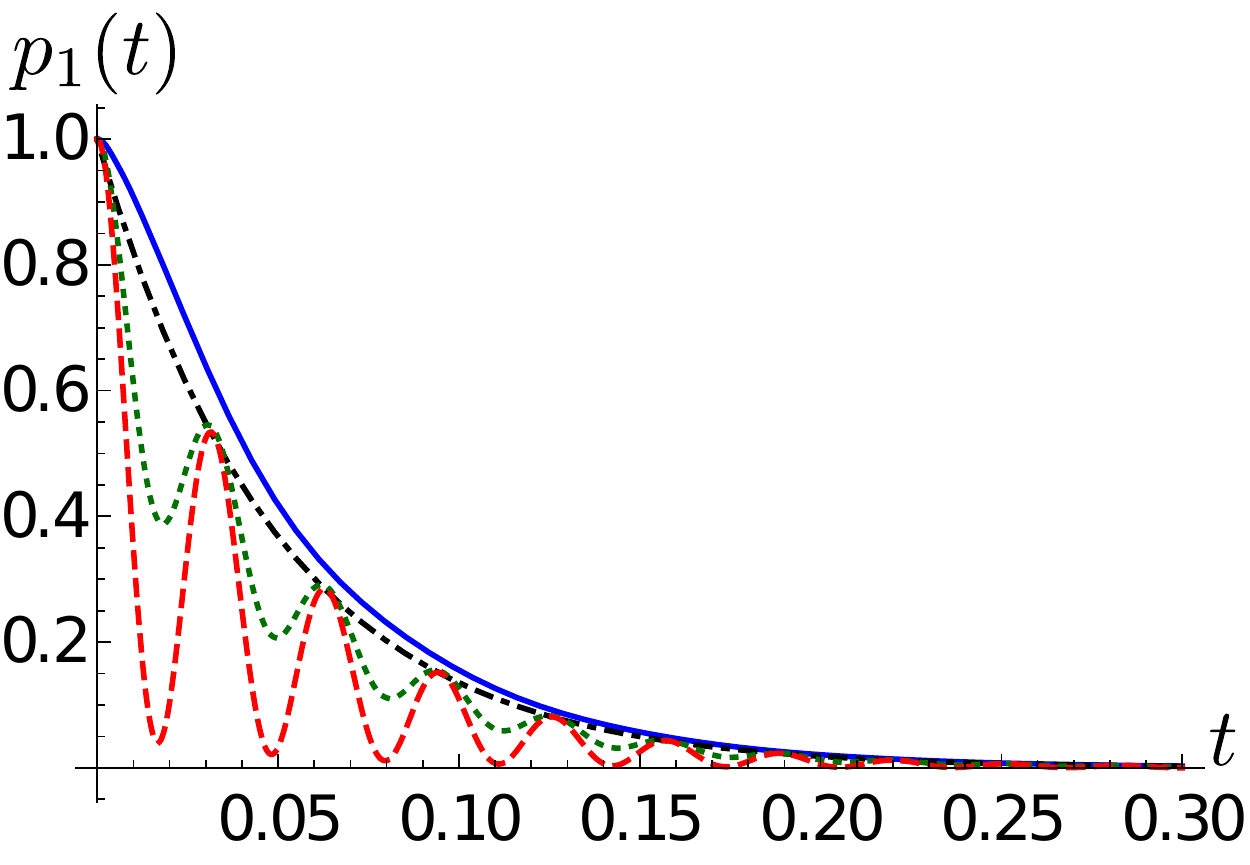}}
\caption{\label{fig:LorentzTempDep} Lorentzian SD. The population $p_1(t)$ of the first TEDOPA chain as a function of time for $T=0$ (blue solid line), $T=77$ (green dotted line) and $T=300$ (red dashed line) for (a) $\gamma=0.001$, (b) $\gamma=1$ and (c) $\gamma=10$. In all plots $\exp(-2 \gamma t)$ is show as a black dot-dashed line as a guide to the eye.}
\end{figure}

In order to provide an insight on how the chain dynamics depends on the temperature, in \Fref{fig:LorentzTempDep} we show the population of the first TEDOPA chain site for different values of $T$. As already observed, the decay rate and the frequency of the population oscillations are independent of $T$, which determines instead the amplitude of such oscillations. This leads us to the conclusion that the oscillation frequency must be determined by the parameter $\Omega$, as expected.

\subsection{Ohmic spectrum} \label{sec:Ohmic}
We now consider spectral densities belonging to the Ohmic family, defined as in Eq.\ref{eq:ohmicSD}. More in particular, we will study the chain dynamics on TEDOPA chains corresponding to the choice $s=0.5, 1$ and $2$, representative, respectively, of sub-Ohmic, Ohmic, and super-Ohmic spectral densities. In all the following examples we will set  $\omega_c  = 100$, and enforce a hard cutoff $\omega_\text{hc} = 10 \omega_c$.

\begin{figure} 
\centering
\subfigure[]{ \includegraphics[width=0.3 \textwidth]{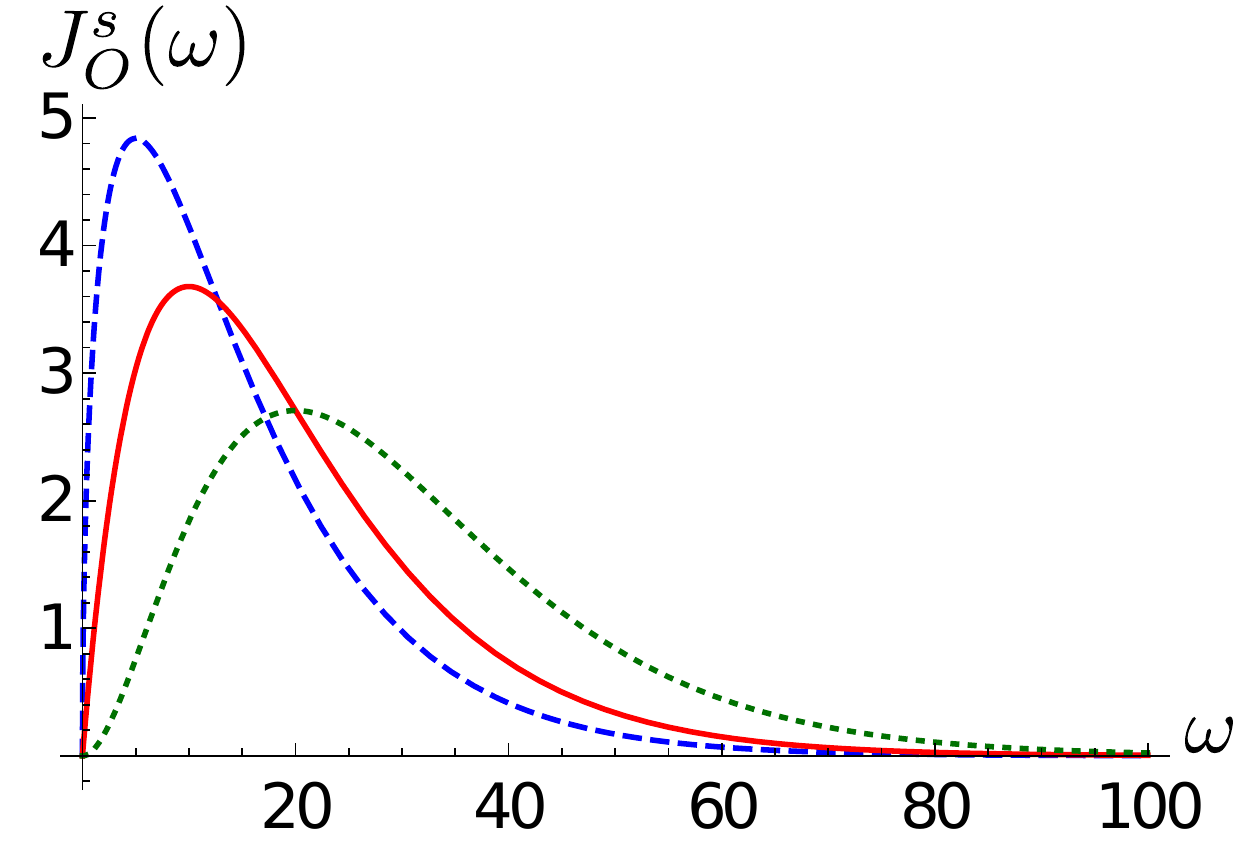}}
\subfigure[]{\includegraphics[width=0.3 \textwidth]{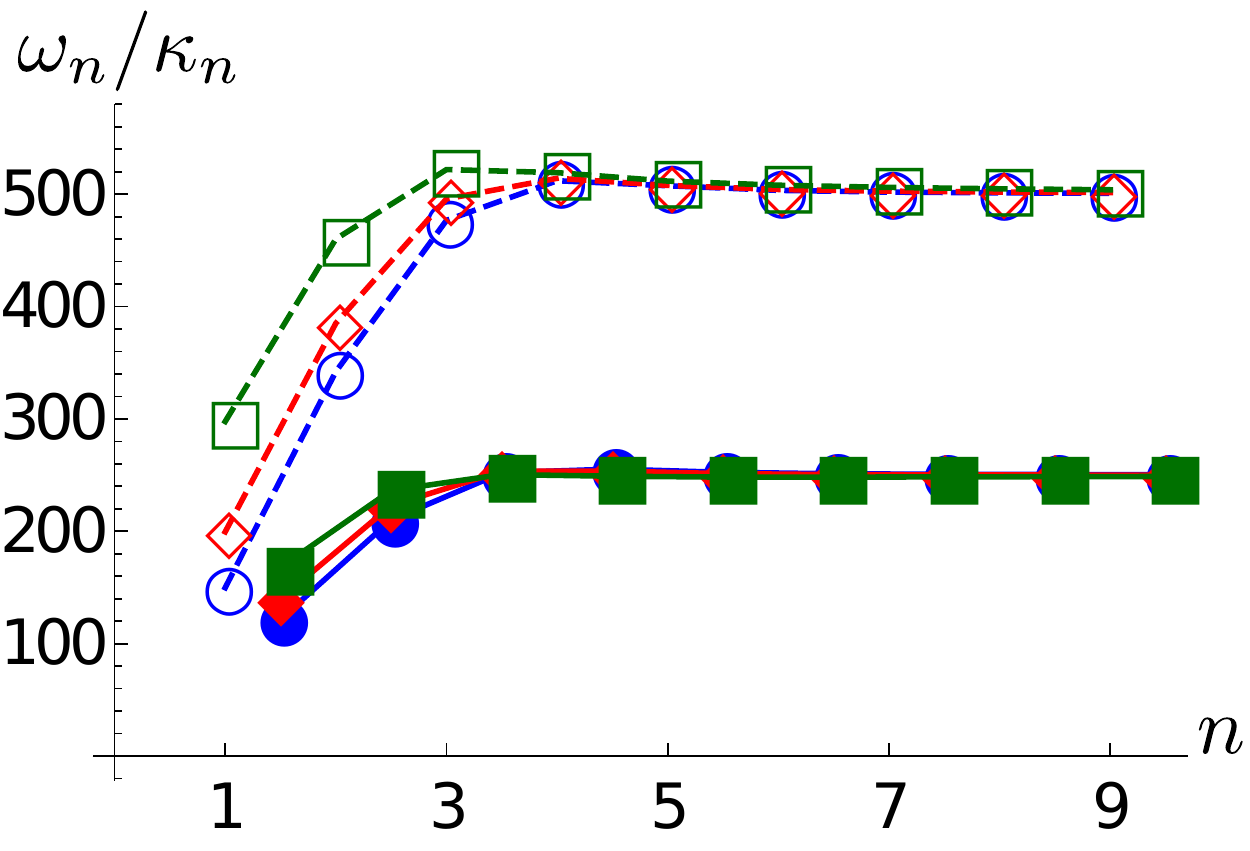}}
\subfigure[]{\includegraphics[width=0.3 \textwidth]{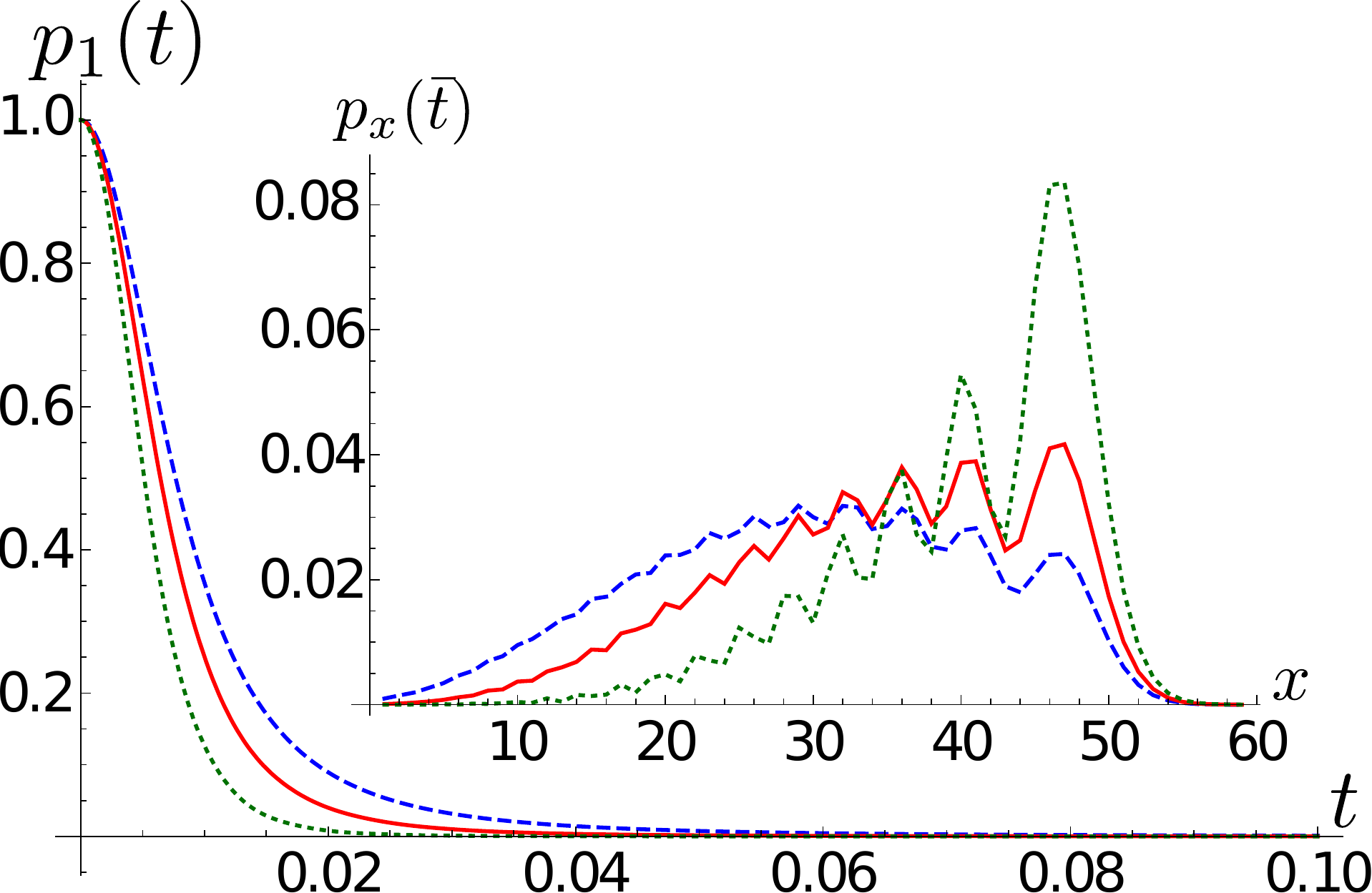}}\\
\subfigure[]{\includegraphics[width=0.3 \textwidth]{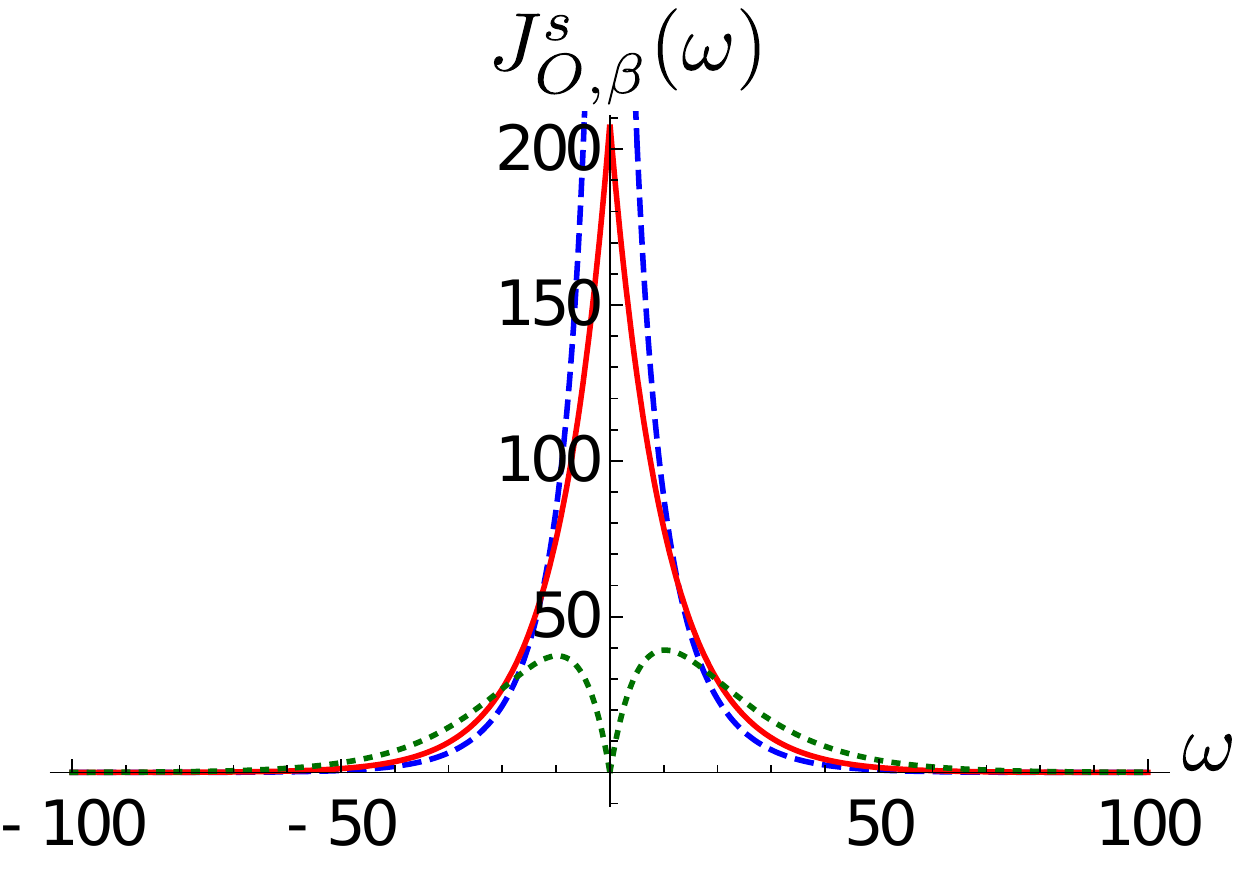}}
\subfigure[]{\includegraphics[width=0.3 \textwidth]{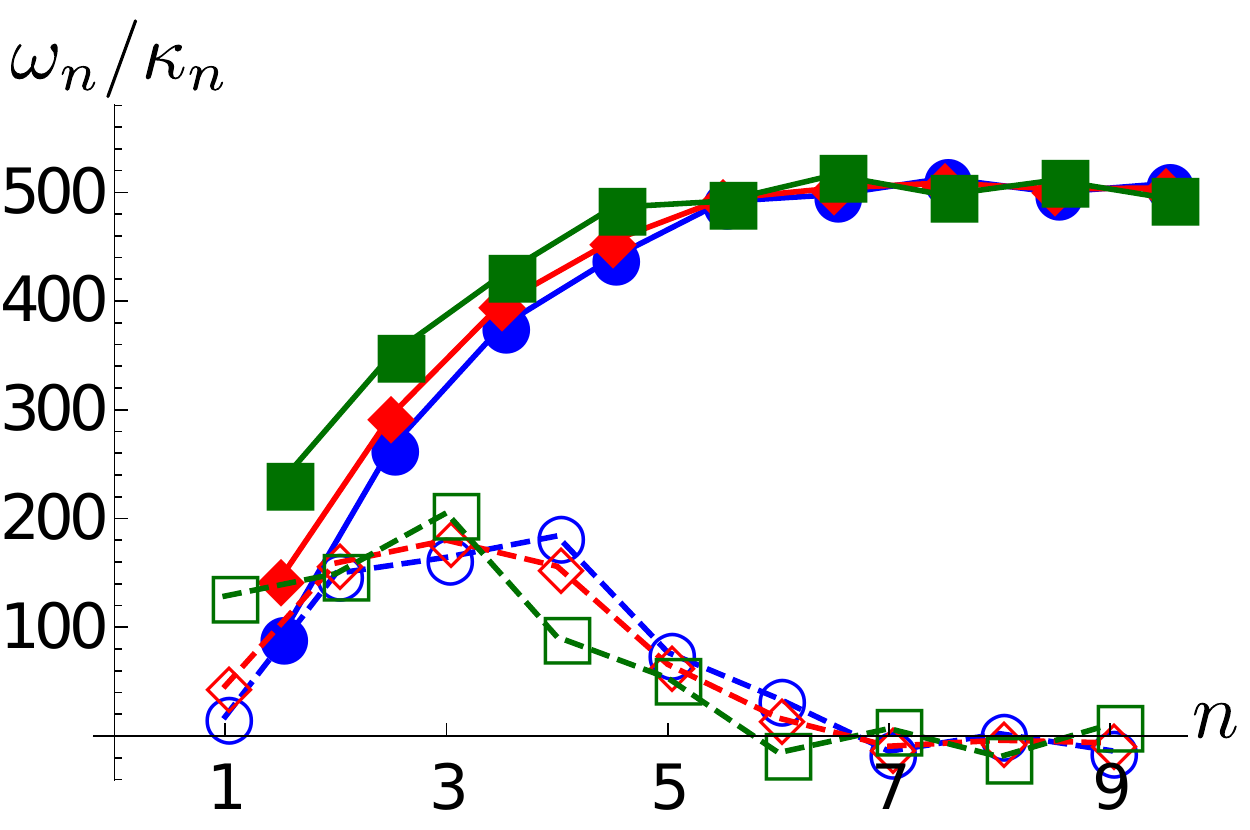}}
\subfigure[]{\includegraphics[width=0.3 \textwidth]{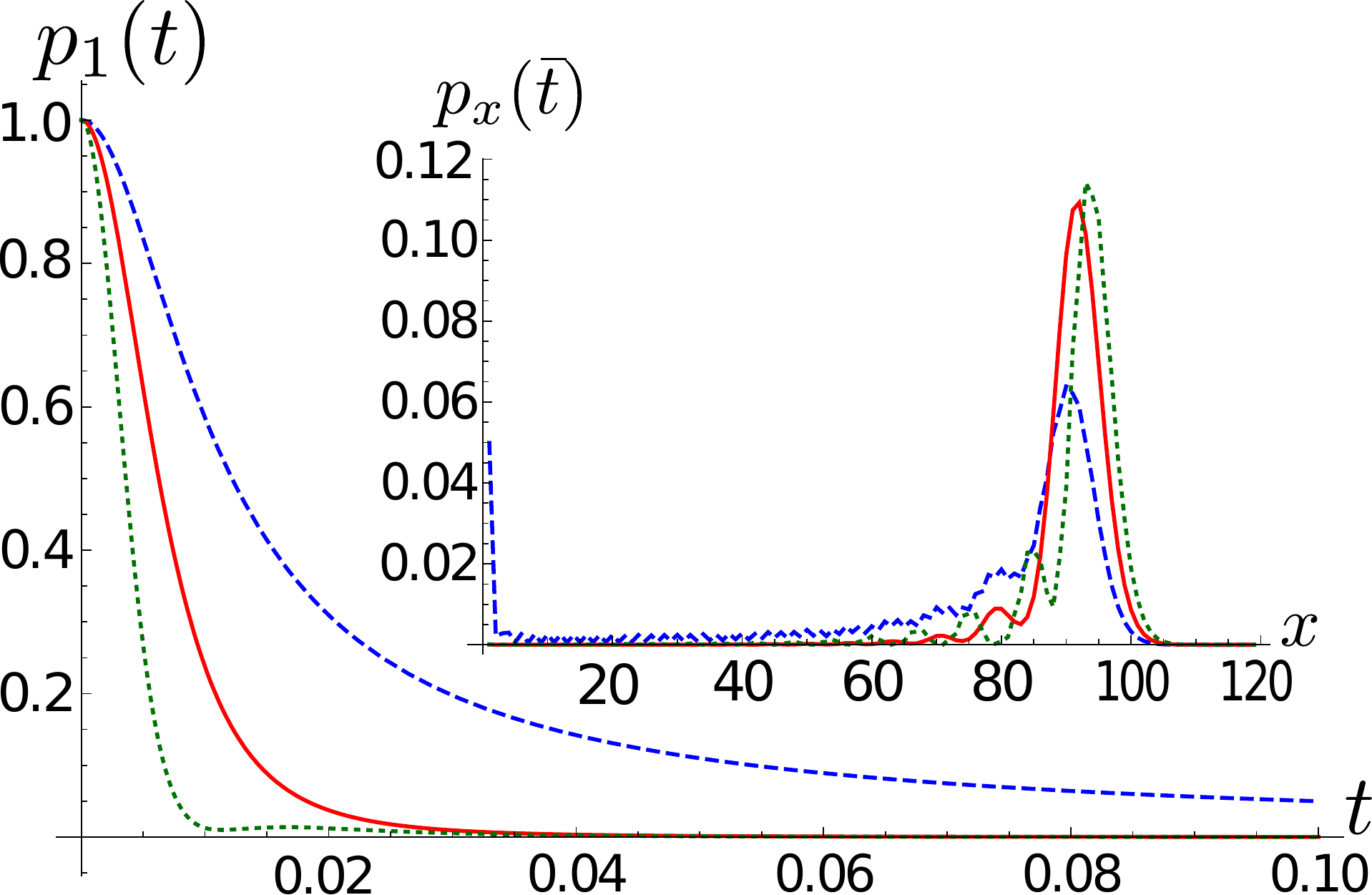}}
\caption{\label{fig:ohmicSDcoeff} Ohmic SD. In all frames $\omega_c=100$, and red markers/solid lines, blue markers/dashed lines, green markers/dotted lines correspond, respectively, to the Ohmic ($s=1$), sub-Ohmic ($s=0.5$) and super-Ohmic ($s=2$) cases.  (a) $T=0$; the spectral density \eref{eq:ohmicSD} for $s=0.5,\ 1, \ 2$. (b) $T=0$; the chain coefficients $\omega_n$ (empty markers), $\kappa_n$ (filled markers). (c) The population of the first chain site as a function of time; in the inset, the populations $p_x(\bar{t})$ for $\bar{t}=0.1$ as a function of $x$. (d) The thermalized SD $J_{O,\beta}^s(\omega)$ for $s=0.5,1,2$ at $T=300$. (e)-(f) Same quantities as frames (b)-(c) for $T=300$.
}
\end{figure}

We start by the $T=0$ case. Frames (a) and (b) of \Fref{fig:ohmicSDcoeff} show Ohmic spectral densities for the selected values of $s$ and the corresponding chain coefficients.  The chain dynamics shows that an excitation leaves its initial location faster in the super-Ohmic case  than in the Ohmic and sub-Ohmic case (see \Fref{fig:ohmicSDcoeff}(c)). This can be justified by the higher coupling coefficient and smaller detuning between the first sites of the TEDOPA chain in the $s=2$ case with respect to the $s=0.5,1$ cases. Moreover, even if the front of the excitation wavepacket travels at the same speed in the three cases, the delocalization degree of the wavepacket is higher in the sub-Ohmic case, while it remains more ``compact'' in the super-Ohmic case, as examplified by the inset of \Fref{fig:ohmicSDcoeff}(c), showing the TEDOPA chain site populations $p_x(t)$ at $t=0.1$. Considered that the chain coefficients for $s=0.5,1,2$ are very close to each other for $n \geq 4$, this difference is explained by the first chain coefficients. Roughly speaking, the higher coupling and smaller detuning between the first chain sites in the $s=2$ case allows for  more compact evolution of the  wavepacket in the momentum space. 

In the high-temperature regime $T=300$, the main features of the chain dynamics are preserved, though with some differences. The decrease of population the first TEDOPA chain oscillator is  still slower in the sub-Ohmic case; for the Ohmic SD, the first site population decay is similar to the $T=0$ case, whereas for the super-Ohmic SD such decay is faster than in the zero temperature scenario (compare frames (c) and (f) of  \Fref{fig:ohmicSDcoeff}). As already discussed before, this behaviour is mainly due to the detuning $|\omega_1 - \omega_2|$ and the coupling strength $\kappa_1$ between the first and second TEDOPA chain oscillators. Interestingly enough, for $s=0.5$ part of the wavepacket remains localized at the first chain site, as shown in the inset of \Fref{fig:ohmicSDcoeff}(f) and, as in the T=0 case discussed above, the wavepacket is more delocalized in the sub-Ohmic case than in the super-Ohmic case, with the Ohmic case lying in between. As a last remark, we observe that, similarly to the finite temperature Lorentzian case, the propagation speed of the wavepacket is about twice as large as in the zero temperature case; as already discussed, this is due to the asymptotic relations \eref{eq:asymcoeff}.


%
\begin{figure} 
\centering
\subfigure[]{ \includegraphics[width=0.3 \textwidth]{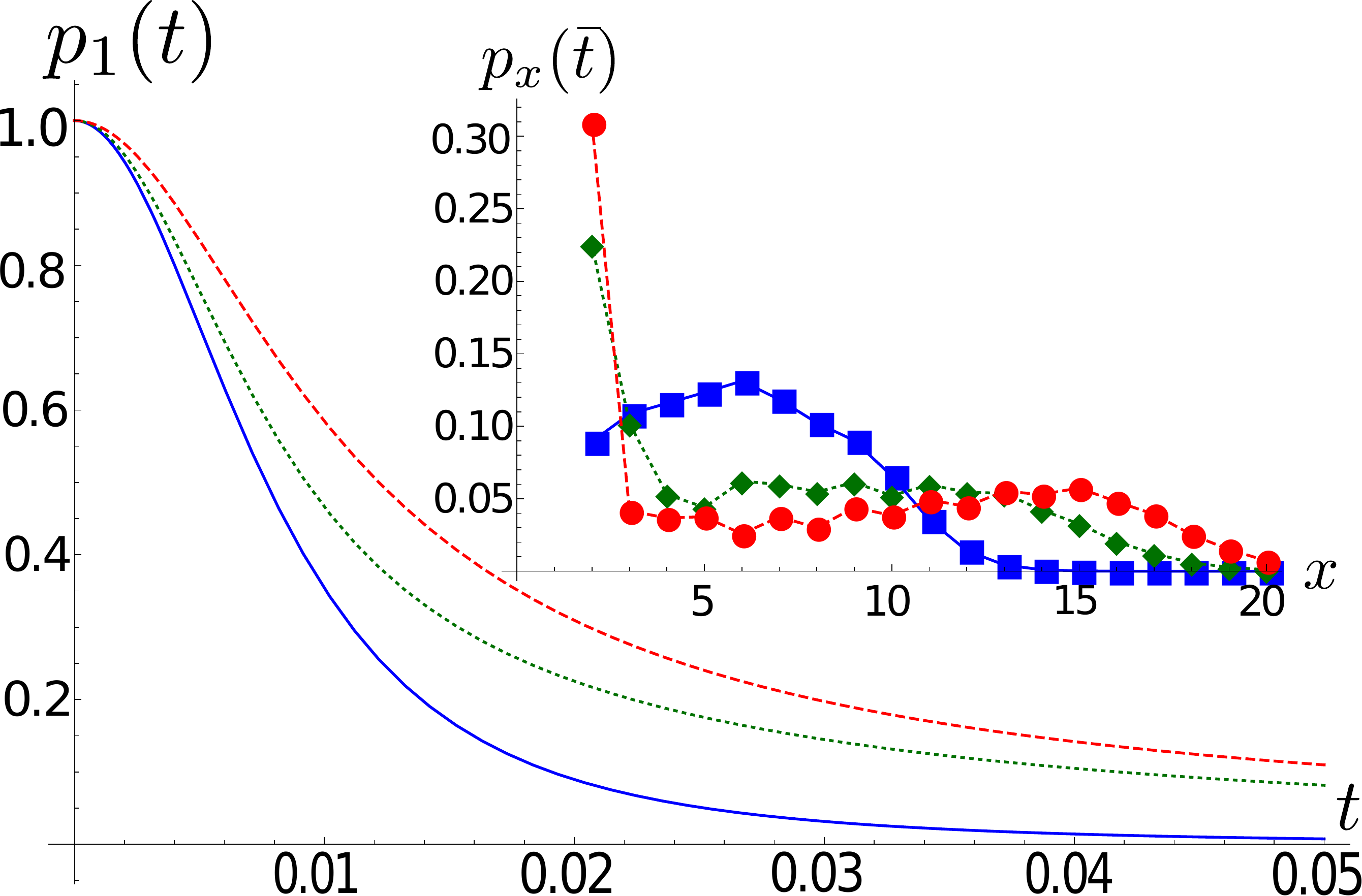}}
\subfigure[]{\includegraphics[width=0.3 \textwidth]{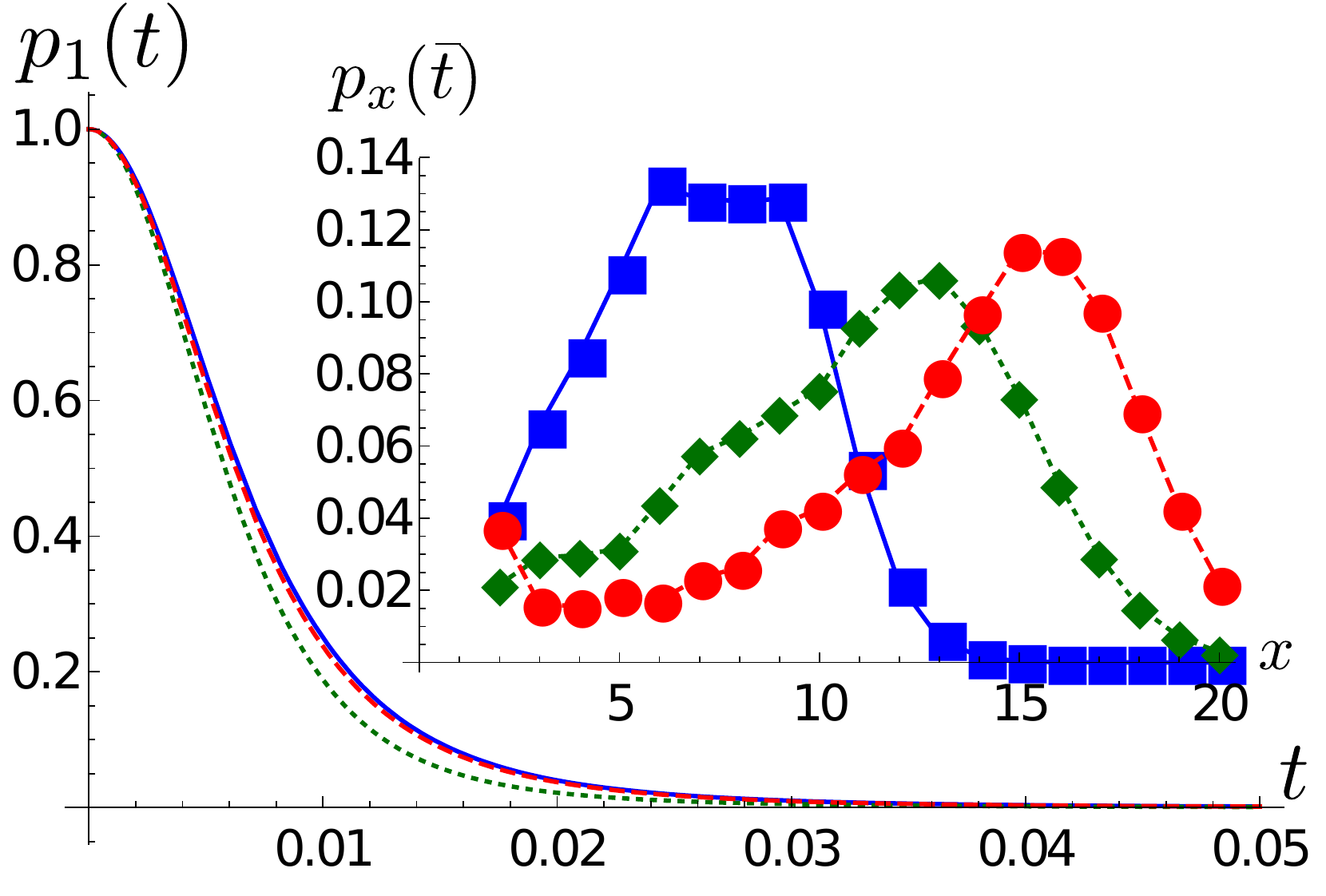}}
\subfigure[]{\includegraphics[width=0.3 \textwidth]{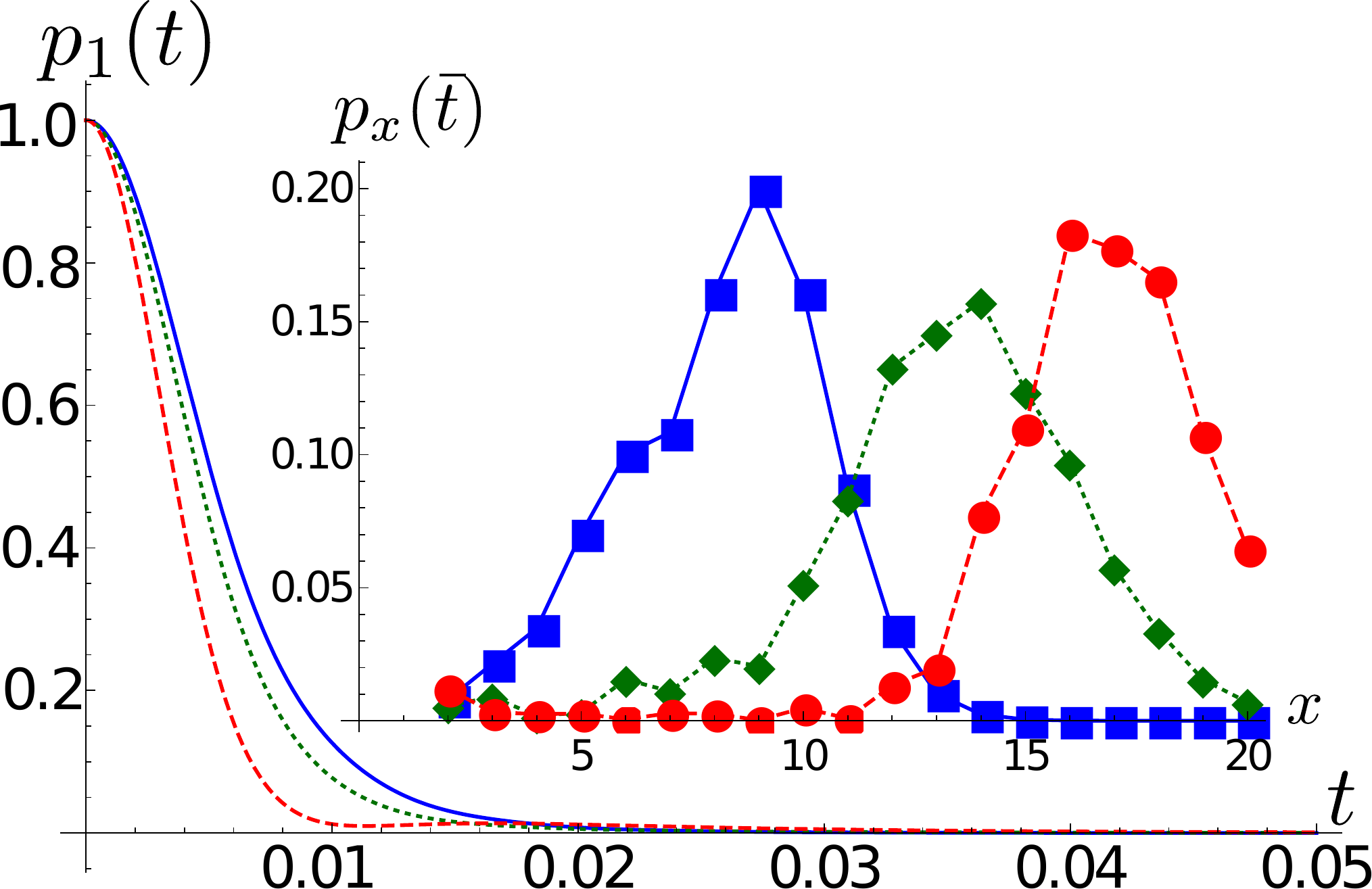}}
\caption{\label{fig:OhmTempDep} Ohmic SD. The population $p_1(t)$ of the first TEDOPA chain as a function of time for $T=0$ (blue solid line), $T=77$ (green dotted line) and $T=300$ (red dashed line) for (a) $s=0.5$, (b) $s=1$ and (c) $s=2$. In the inset of all frames, the population $p_x(\bar{t})$ at $\bar{t}=0.02$.}
\end{figure}

Figure \ref{fig:OhmTempDep} provides more details.  As we did for the Lorentzian SD case, we now inspect the dynamics of the first TEDOPA chain population for the three considered spectral densities at different temperatures. It clearly shows that, while for the Ohmic spectral density such population is only slightly affected by the value of  $T$,  the temperature has opposite effects on super- and sub-Ohmic SDs. As a matter of fact, whereas for the sub-Ohmic case, an increasing temperature leads to a slower decrease of the first site population, in the for $s=2$ the first site empties at a rate which is directly proportional to the temperature. The snapshots on the populations $p_x(t)$ for $t=0.02$ in the insets of frames (a)-(c) of the same figure, allows us to better appreciate the partial trapping at finite temperature of the wavepacket at the first TEDOPA chain site and the more pronounced spreading of the wavepacket in the $s=0.5$ case.

\section{Full dynamics} \label{sec:fulldyn}
So far we focused our analysis on the dynamics of a single excitation moving along TEDOPA chains. This allowed us to isolate the main features of such dynamics for representative spectral densities and to investigate the dependence of the kinematic properties of TEDOPA chains on the specific form of the SD and on the temperature. Clearly enough, the single excitation subspace we restricted ourselves to is not suited to describe the chain dynamics in the presence of a system interacting with the environment. As a matter of fact, the interaction with the system will dynamically inject in (and subtract from) the chain excitation, at a rate that depends, among other things, on the system-environment coupling strength.

In this section, therefore, we extend our analysis by considering a two-level system interacting with a bosonic environment described by either Lorentzian or Ohmic spectral densities. Given the spectral density, the spin-boson model is fully specified once the system and the system-environment interaction Hamiltonian are fixed. In what follows, we  specialize the Hamiltonian \eref{eq:totHam} to 
\begin{align}
H_S &= \Delta \sigma_x  \\
A_S &= \frac{1+\sigma_z}{2} \\
O_\omega &= X_\omega = (a_\omega + a_\omega^\dagger),
\end{align}
with $\sigma_x,\sigma_z$ Pauli matrices, describing, for example, an homo-dimer interacting with a vibronic environment \cite{plenio13}. The resulting dynamics is therefore not a pure dephasing dynamics, and is representative of the class of physical systems for which numerically exact approaches are required. Considered that the interaction term does not change the system's populations but affects only its coherences, we will initialize the system to the state $\ket{+} = 1/\sqrt{2} (1,1)^T$, namely the eigenstate of $\sigma_x$ belonging to the eigenvalue $+1$, representative of the maximally coherent states in the $\sigma_z$ basis. The initial state of the environment will be instead a thermal state \eref{eq:thermalEnv} at temperature $T$. In the following examples we will set $\Delta=70 \text{cm}^{-1}$, and tune the parameter $\lambda$ of eqs. \eref{eq:asymmLorentz} and \eref{eq:ohmicSD} so that the system-TEDOPA chain coupling $\kappa_0$ (see \eref{eq:overall}) is the same at $T=0$ for all the considered spectral densities. More precisely, by definition, the $k_0$ coefficient of the Ohmic spectral density is independent of $s$ so that, in the Ohmic cases, we set $\lambda=1$; for Lorentzian spectral densities we set to $\lambda=60$.

Before presenting our results it is important to remark that we are not so much interested in the reduced dynamics of the system, but rather on the TEDOPA chain dynamics in the presence of an interaction with the open system. In particular, we will try to understand which of the features discussed in the preceding section persist in the presence of an interaction with the system. To this end we will use the average occupation number 
\begin{equation}
    n_k(t) = \Tr(b_k^\dagger b_k \rho_C(t))
\end{equation}
of the $k$-th chain oscillator where $\rho_C(t)$ is the system+chain state at time $t$ determined via TEDOPA simulation. 

\begin{figure} 
\centering
\subfigure[]{ \includegraphics[width=0.3 \textwidth]{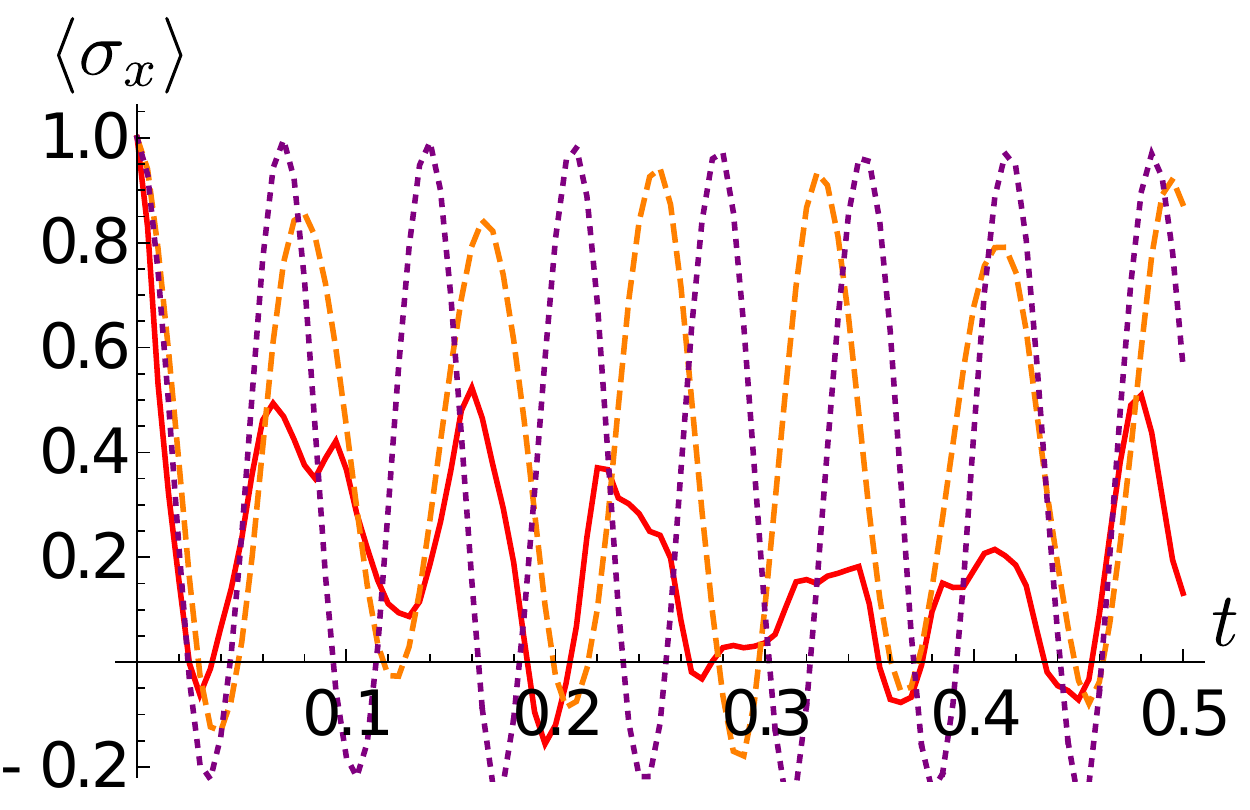}}
\subfigure[]{\includegraphics[width=0.3 \textwidth]{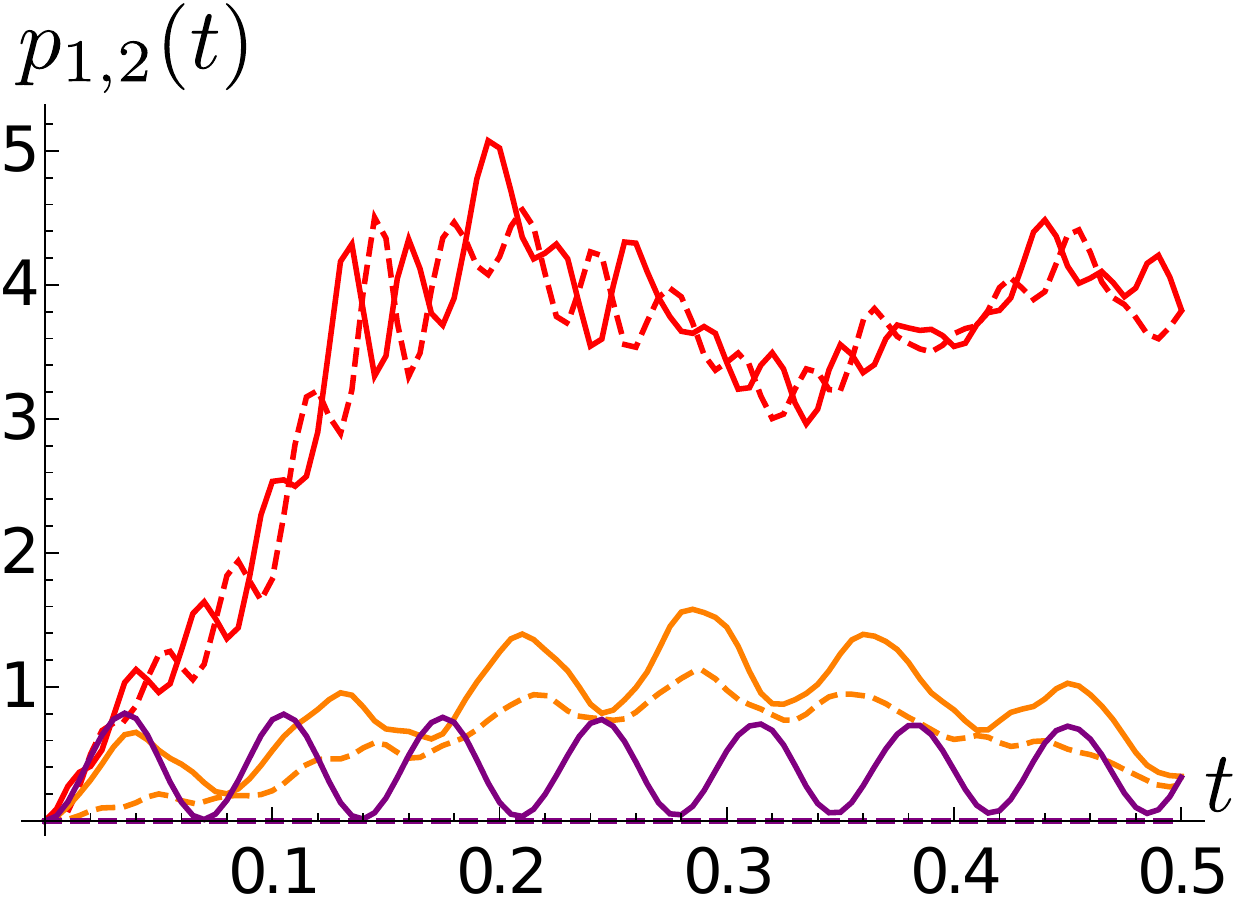}}
\subfigure[]{ \includegraphics[width=0.3 \textwidth]{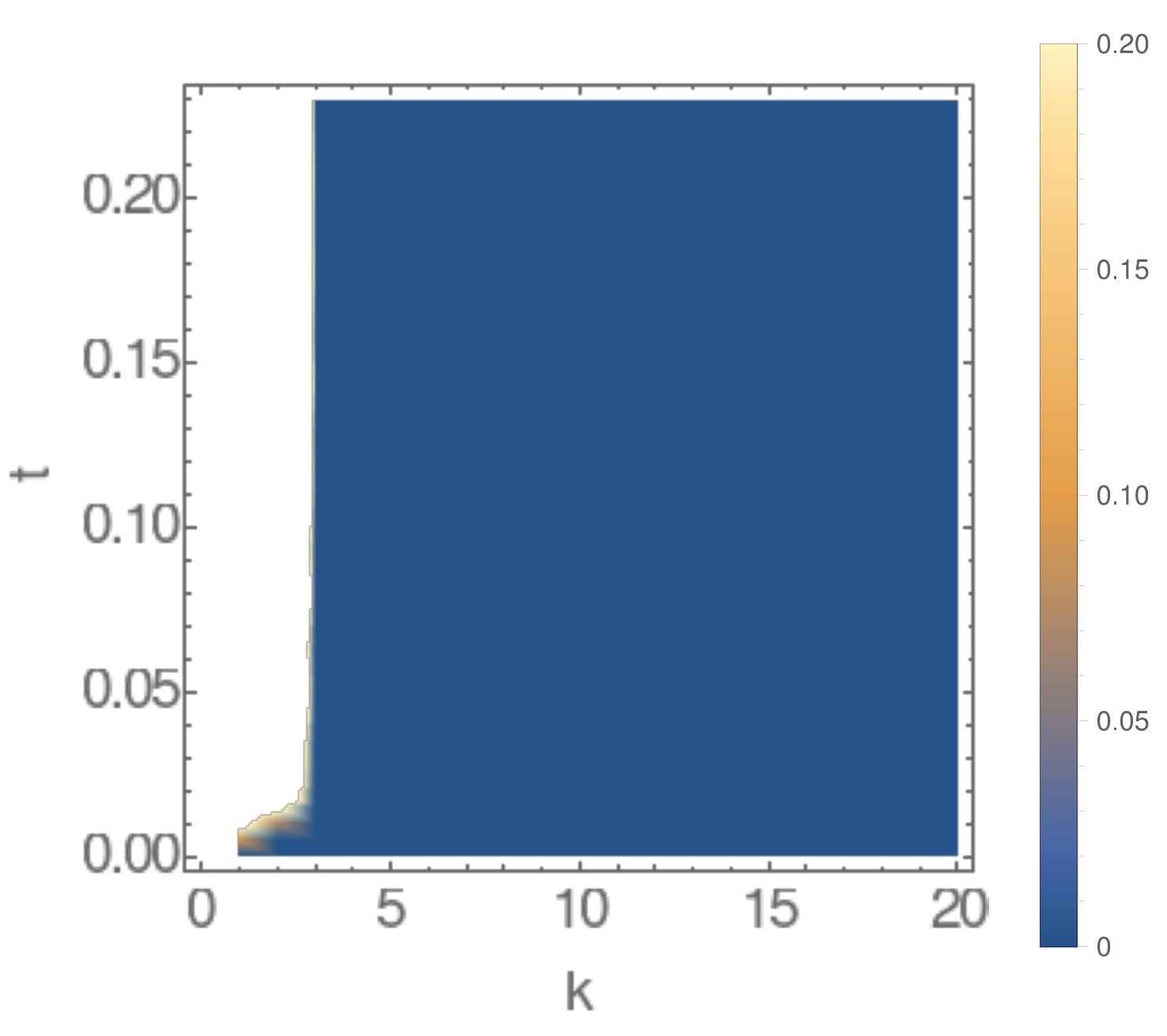}}
\caption{\label{fig:LorentzFullg001} Lorentzian SD, full dynamics. $\gamma=0.001$ (a) The expectation of $\sigma_x$ as a function of time for $T=0$ (purple dotted line) $T=77$ (orange dashed line) and $T=300$ (solid red line). (b) The average occupation number $p_{1,2}(t)$ of the first (solid lines) and the second (dashed line) TEDOPA chain sites for $T=0$ (purple) $T=77$ (orange) and $T=300$ (red). (c) The average occupation number of the chain sites $k, k=1,2,\ldots,20$ as a function of time for $T=300$.}.
\end{figure}

We first discuss the chain dynamics for Lorentzian spectral densities. The $\gamma=0.001$ case is still paradigmatic. At $T=0$ only the first TEDOPA chain oscillator is essentially involved in the dynamics. By comparing the purple lines in frames (a) and (b) of \Fref{fig:LorentzFullg001}, we can clearly see the beatings between the system and the first TEDOPA chain site. For $T>0$ a the second TEDOPA chain mode enters into play. The average occupation number $n_{1,2}(t)$ of the first two chain sites depend on the temperature. Interestingly enough, in the high ($T=300$) temperature regime the both $n_1(t)$ and $n_2(t)$ present small and fast out of phase oscillations, imprinting on the system dynamics a much more erratic dynamics than the  $T=77$ environment, for which such oscillations are slower and almost in phase. 

\begin{figure} 
\centering\subfigure[]{ \includegraphics[width=0.3 \textwidth]{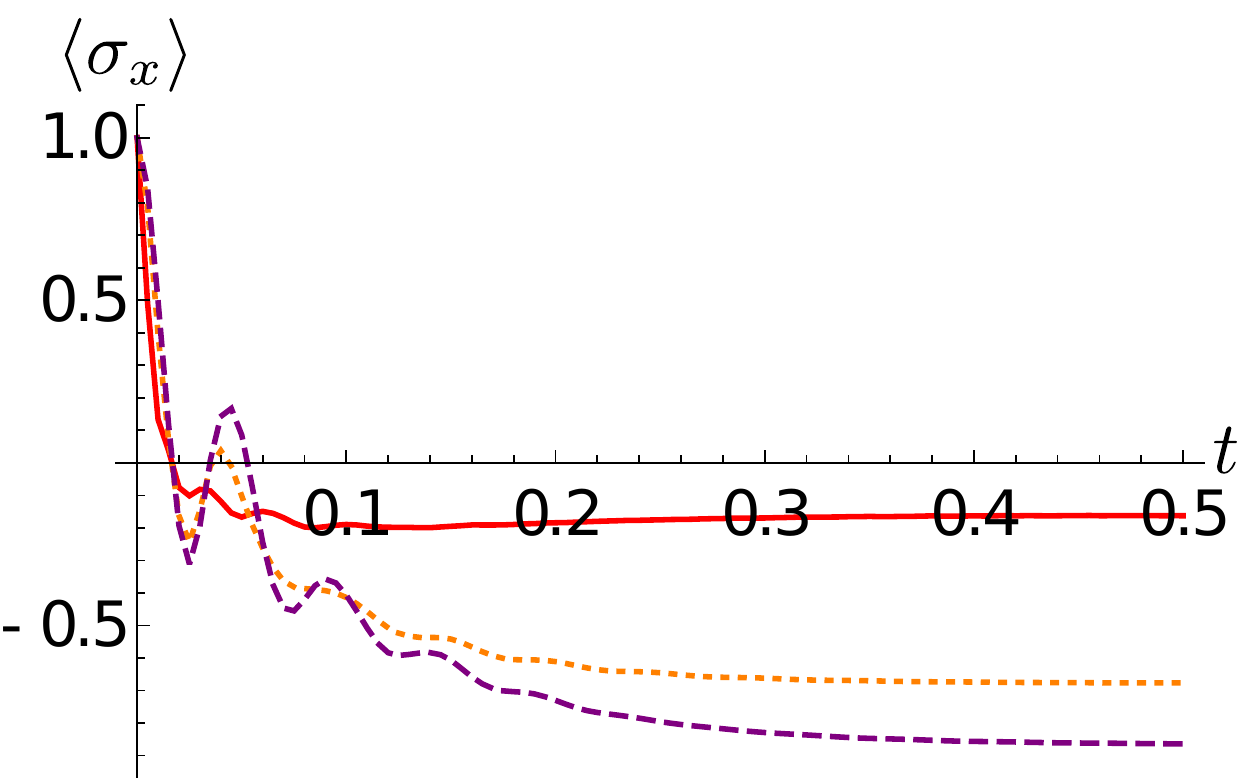}}
\subfigure[]{\includegraphics[width=0.3 \textwidth]{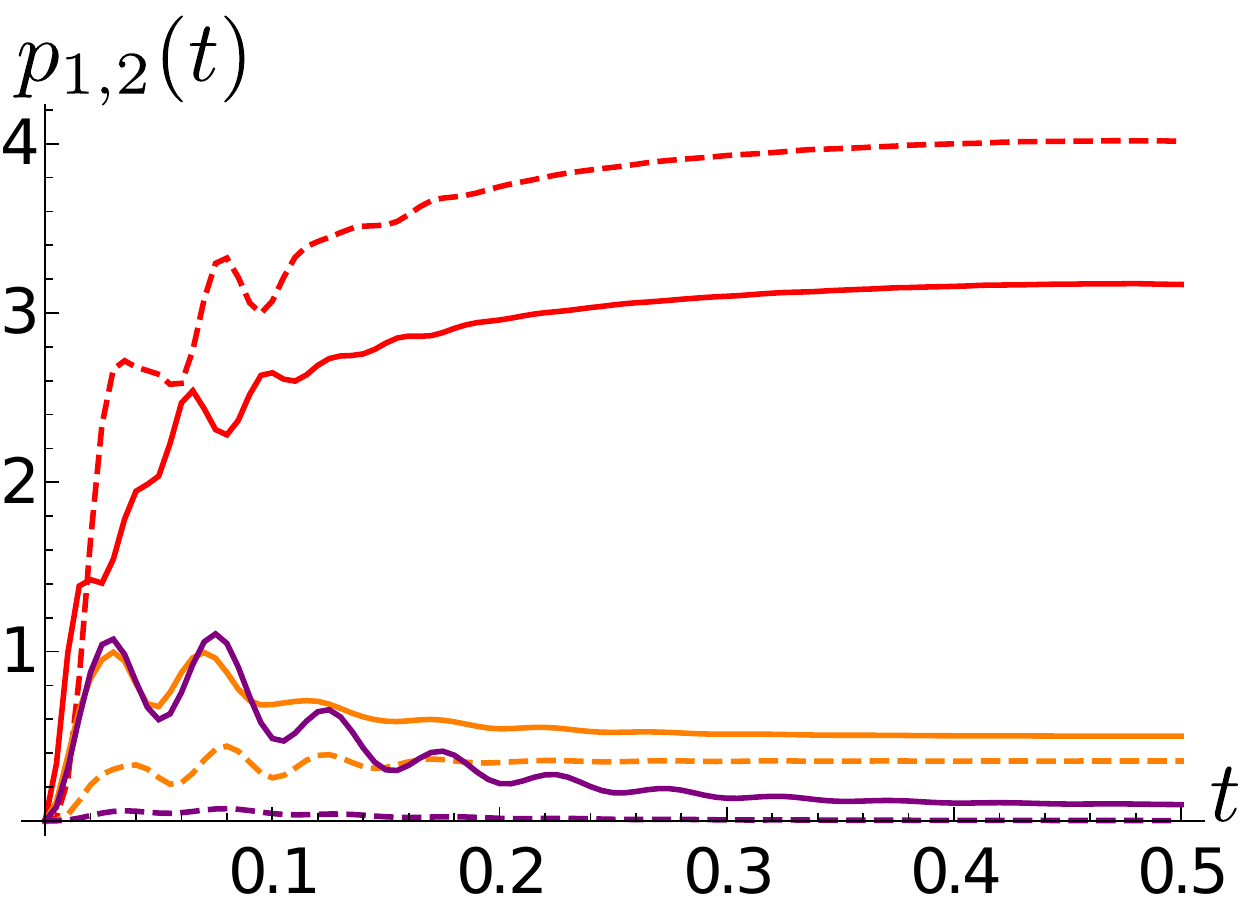}}
\subfigure[]{ \includegraphics[width=0.3 \textwidth]{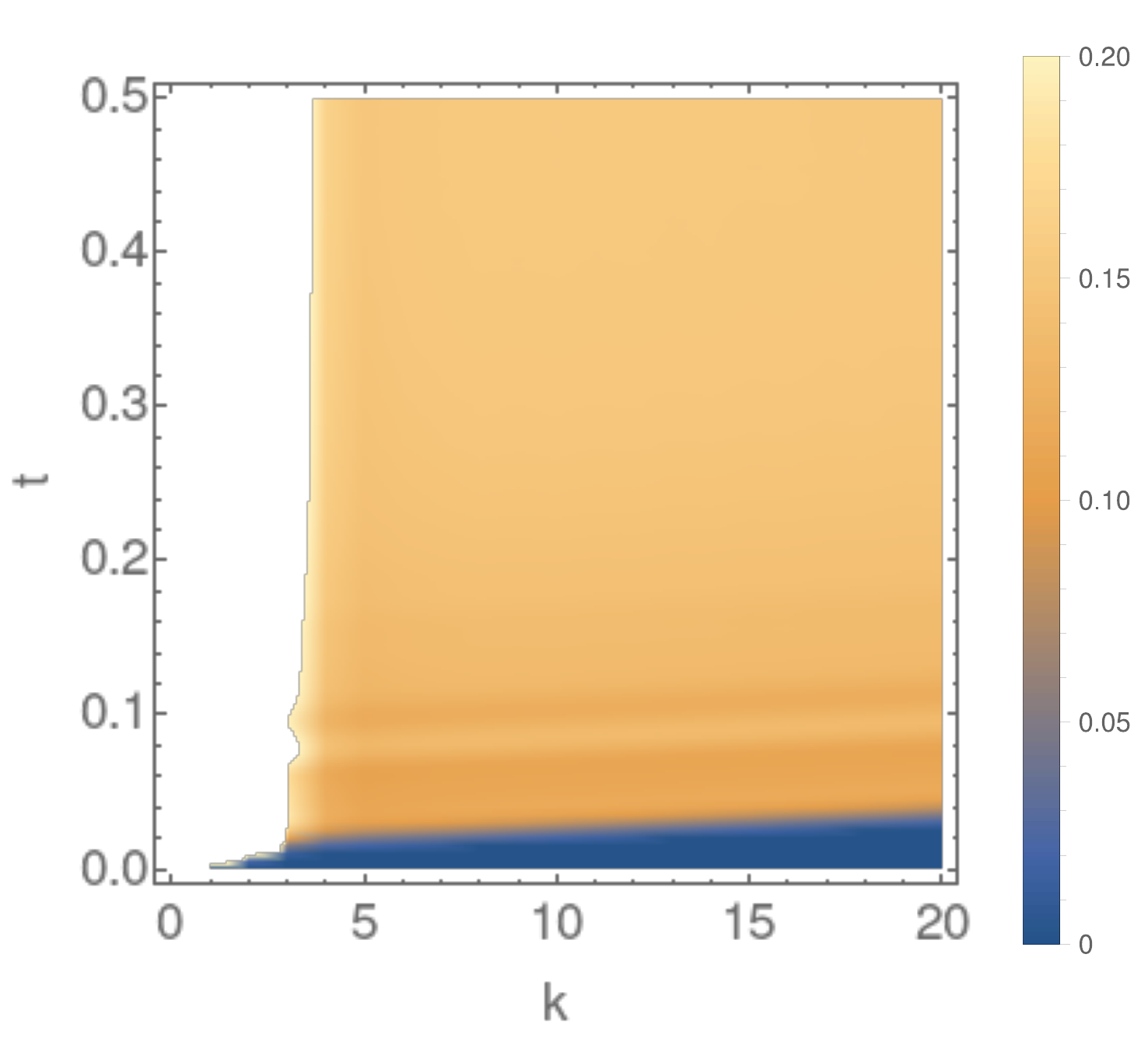}}

\caption{\label{fig:LorentzFullg10} Lorentzian SD, full dynamics.$\gamma=10$. Same quantities and line styles as in \Fref{fig:LorentzFullg001}}.
\end{figure}

Figure \Fref{fig:LorentzFullg10} shows instead the system and chain dynamics for $\gamma=10$. Analogously to the $\gamma=0.001$ the average occupation  of the first two TEDOPA chain sites is temperature dependent. The larger value of $\gamma$ implies that, loosely speaking, more environmental modes are interacting with the system. While the first two sites are still the highest occupied ones, some excitations can percolate to the right part of the chain, as we already observed in the chain dynamics analysis of the previous section (see \Fref{fig:LorentzFT}(f)). Since the system-TEDOPA chain coupling is about the same for the two considered values of $\gamma$, it is such percolation responsible for the faster  relaxation of the system.


\begin{figure} 
\centering\subfigure[]{ \includegraphics[width=0.4 \textwidth]{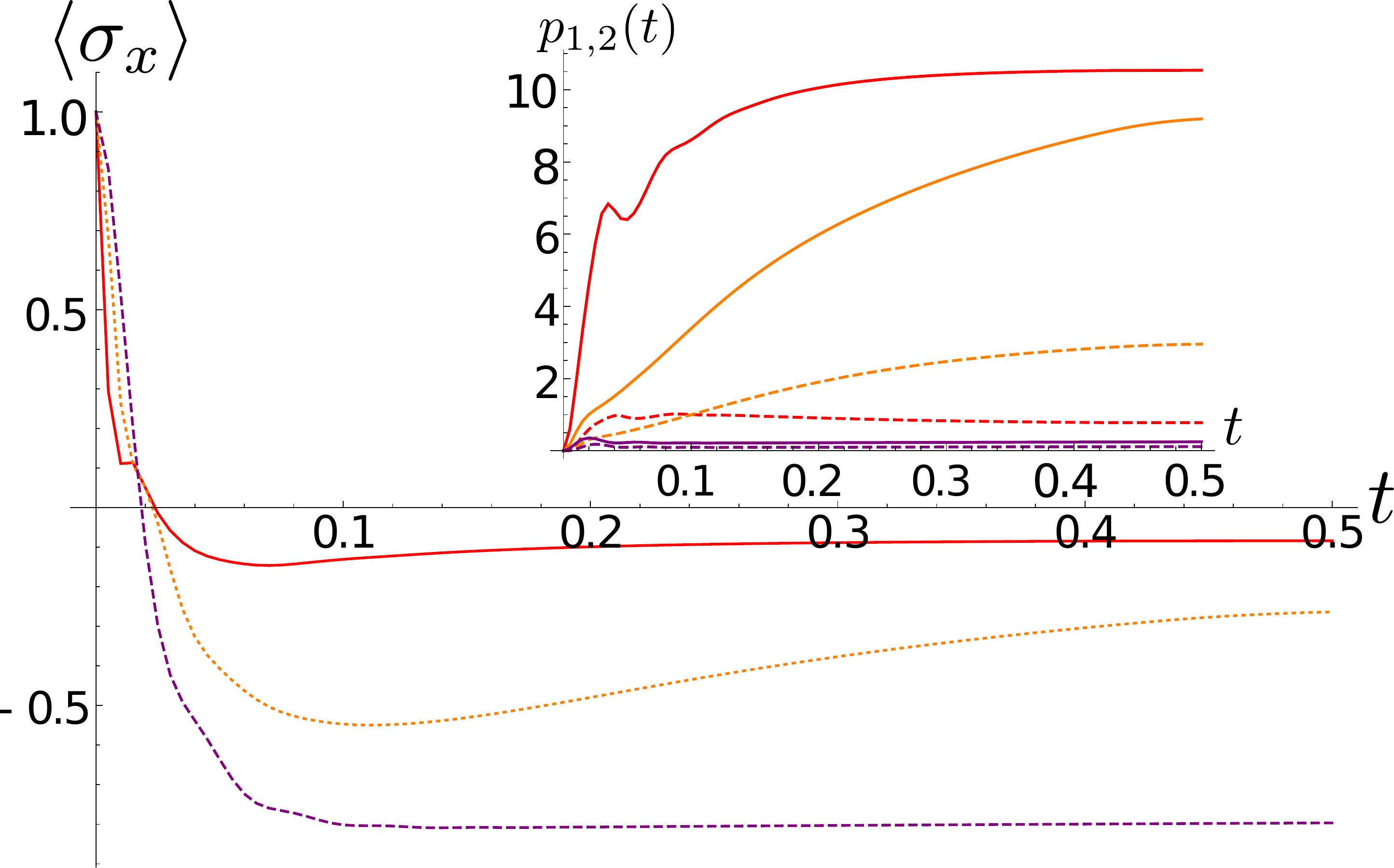}}
\subfigure[]{\includegraphics[width=0.25 \textwidth]{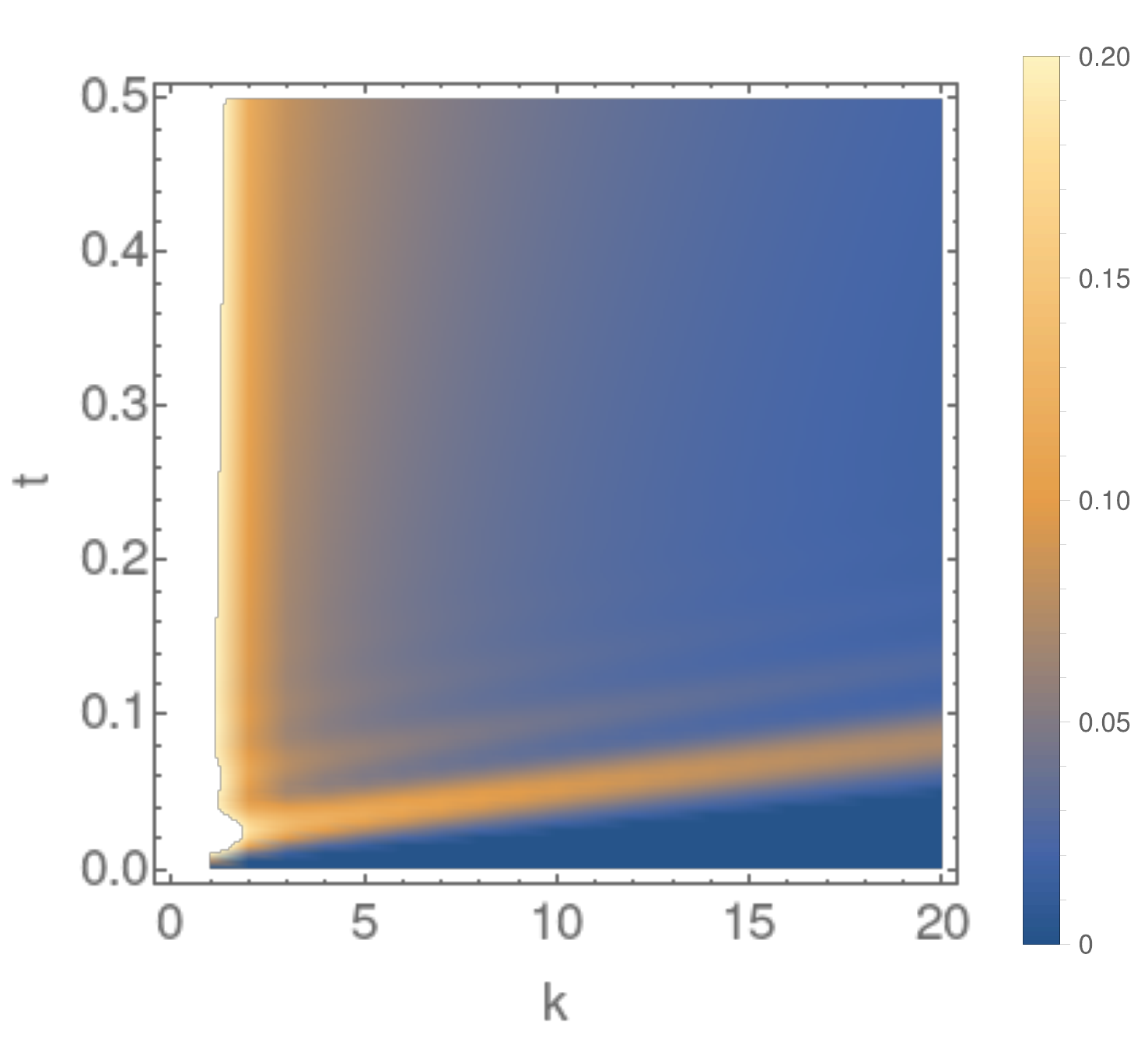}}
\subfigure[]{ \includegraphics[width=0.25 \textwidth]{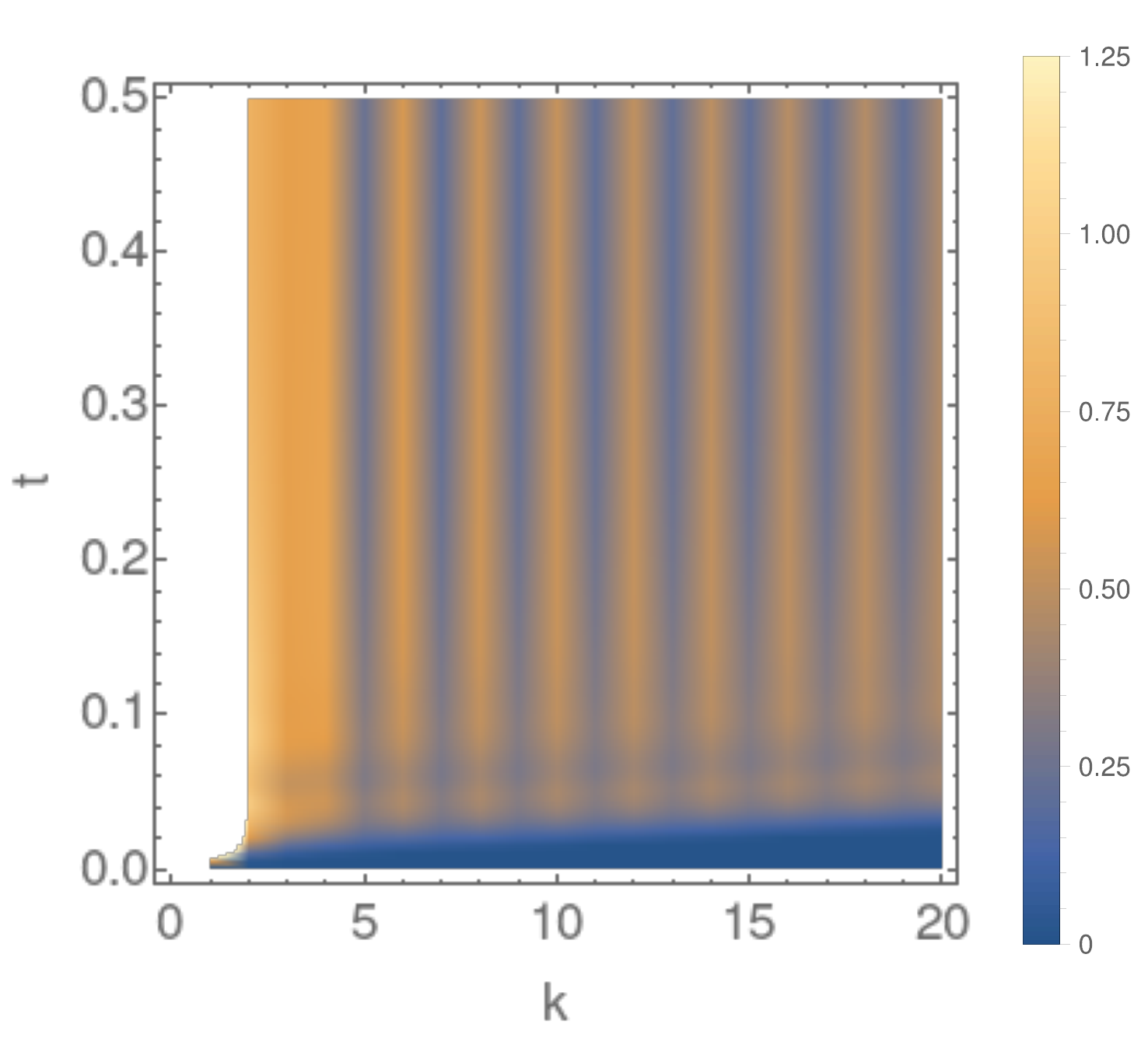}}

\caption{\label{fig:OhmicFulls05} Sub-Ohmic SD ($s=0.5$). (a) The expectation of $\sigma_x$ at different temperatures as a function of time (same line styles as in \Fref{fig:LorentzFullg001}(a)); in the inset, the average occupation number of the first and the second TEDOPA chain oscillators (same line styles as in \Fref{fig:LorentzFullg001}(b)). (b) The average occupation number of the chain sites $k$, for $k=1,2,\ldots,20$ as a function of time at $T=0$. (c) Same quantities as in frame (b) for $T=300$. }
\end{figure}

\begin{figure} 
\centering\subfigure[]{ \includegraphics[width=0.4 \textwidth]{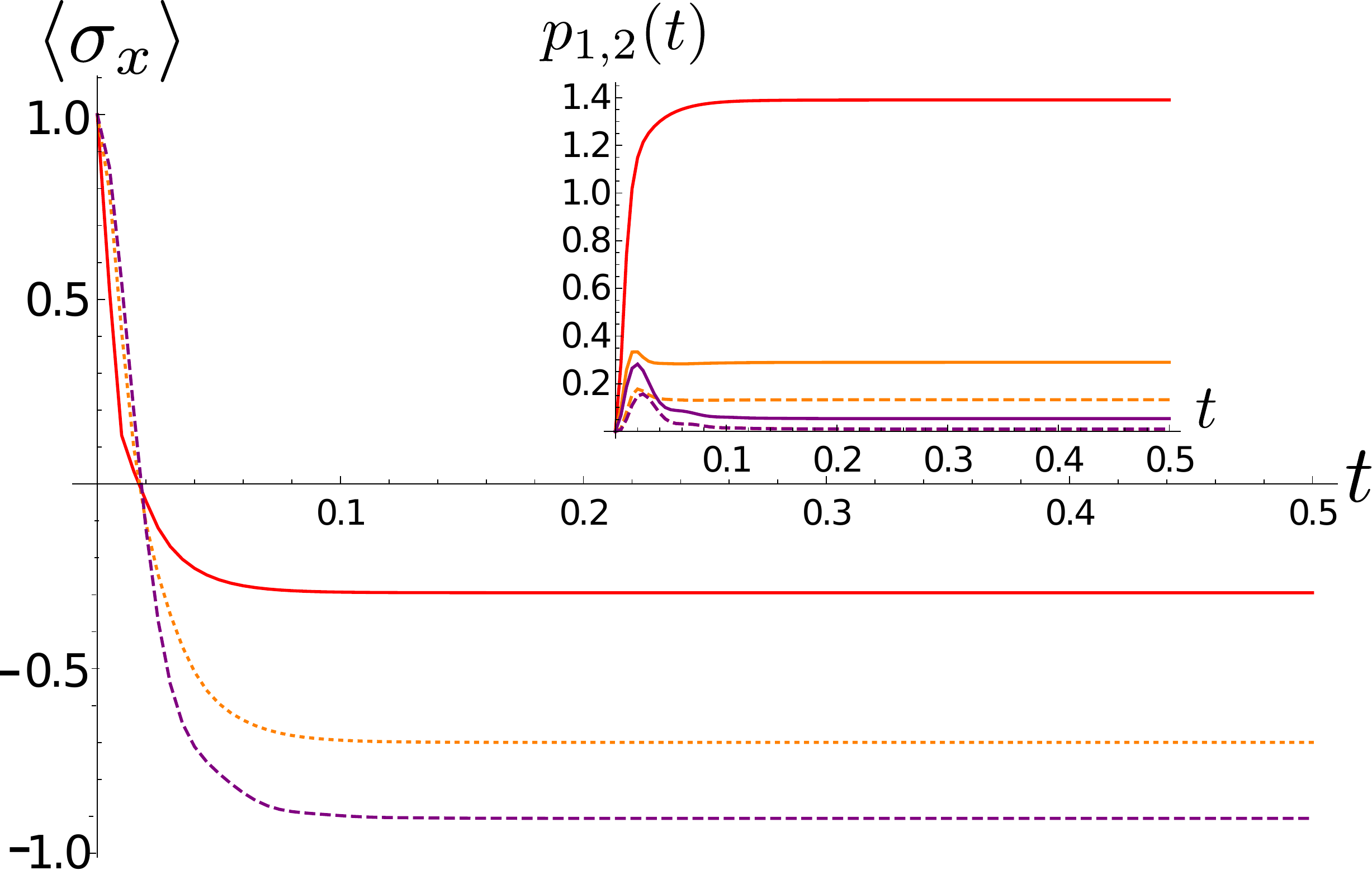}}
\subfigure[]{\includegraphics[width=0.25 \textwidth]{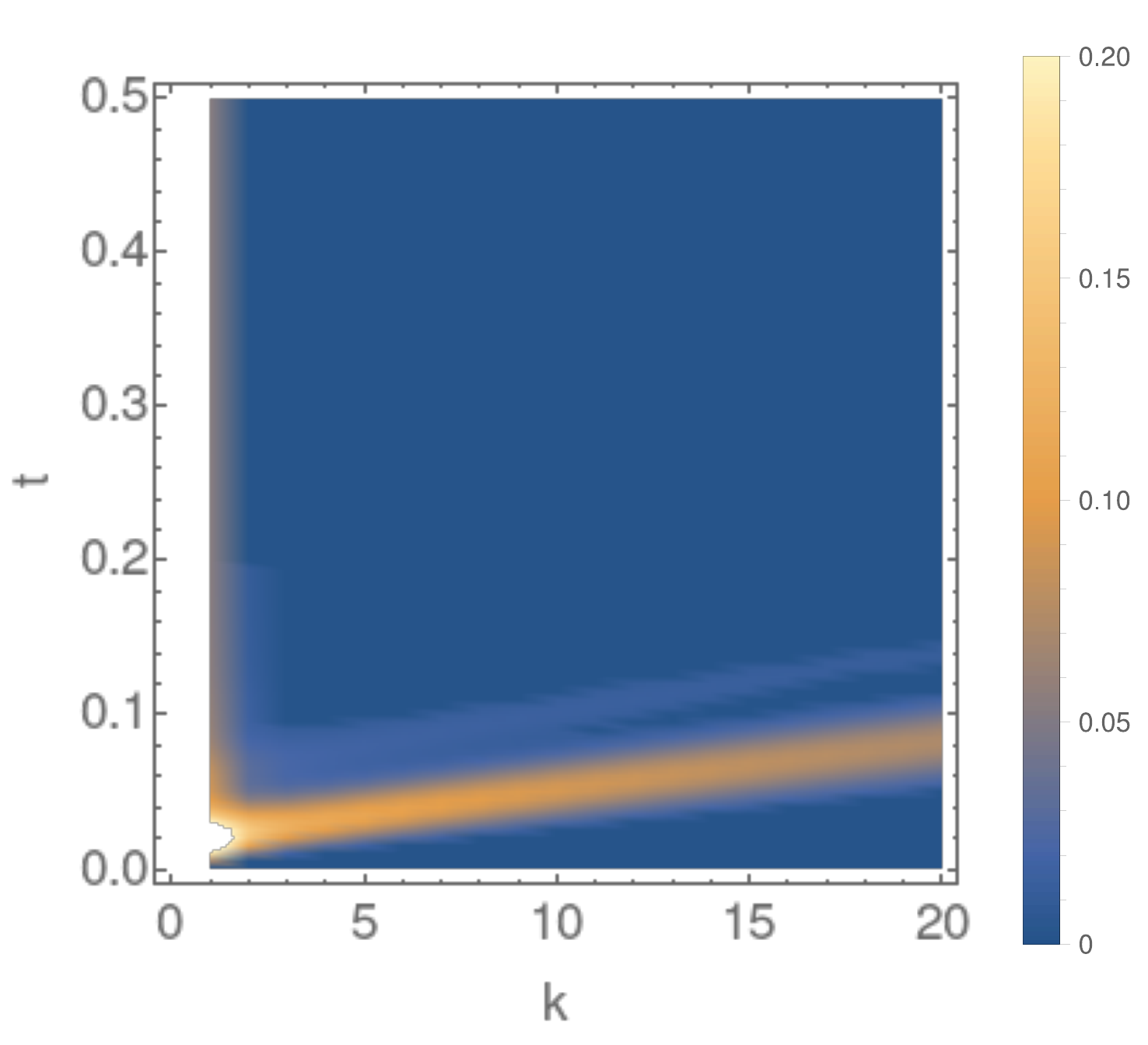}}
\subfigure[]{ \includegraphics[width=0.25 \textwidth]{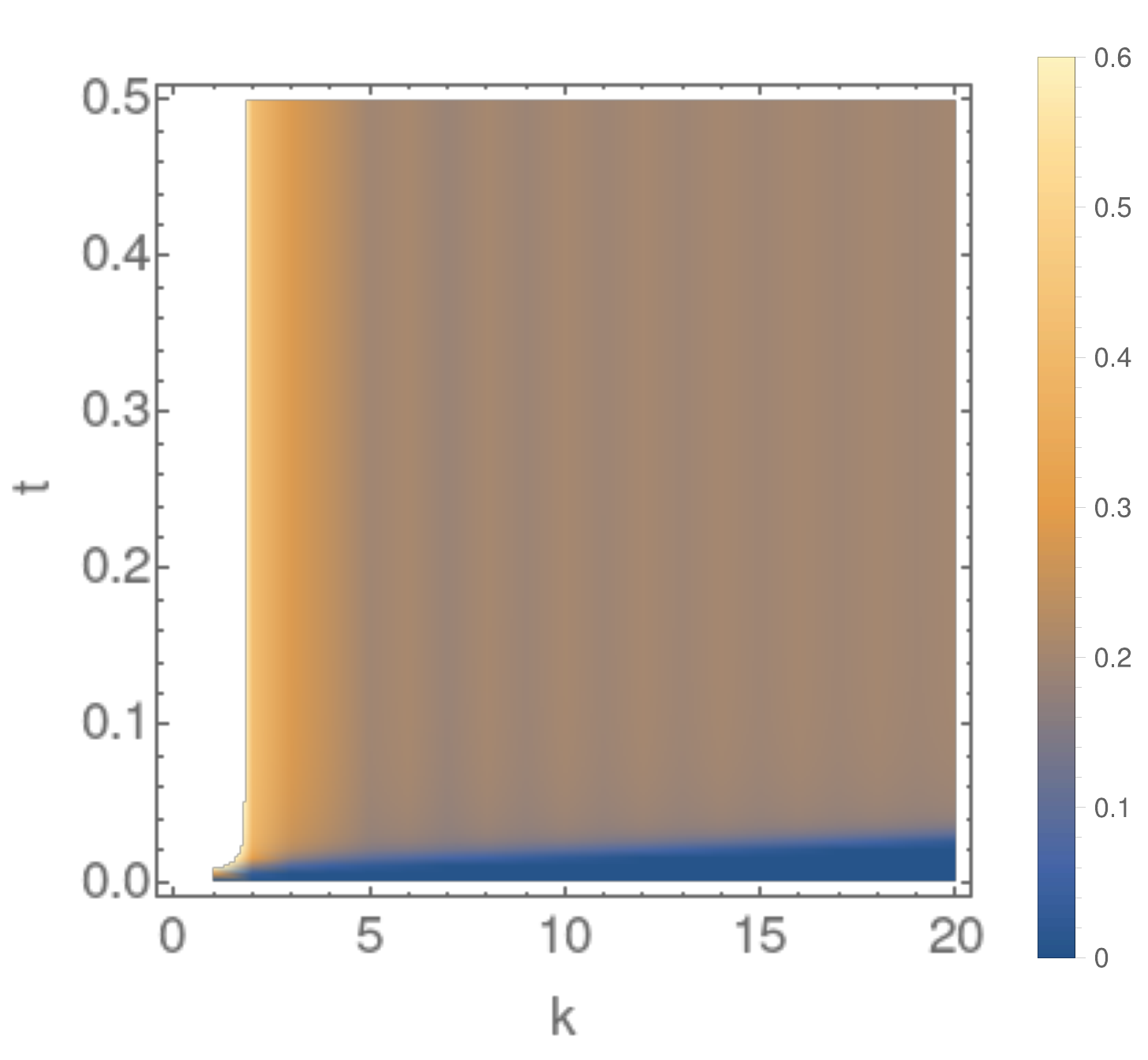}}

\caption{\label{fig:OhmicFulls1} Ohmic SD ($s=1$). Same quantities and line styles as in \Fref{fig:OhmicFulls05}.}
\end{figure}

\begin{figure} 
\centering\subfigure[]{ \includegraphics[width=0.4 \textwidth]{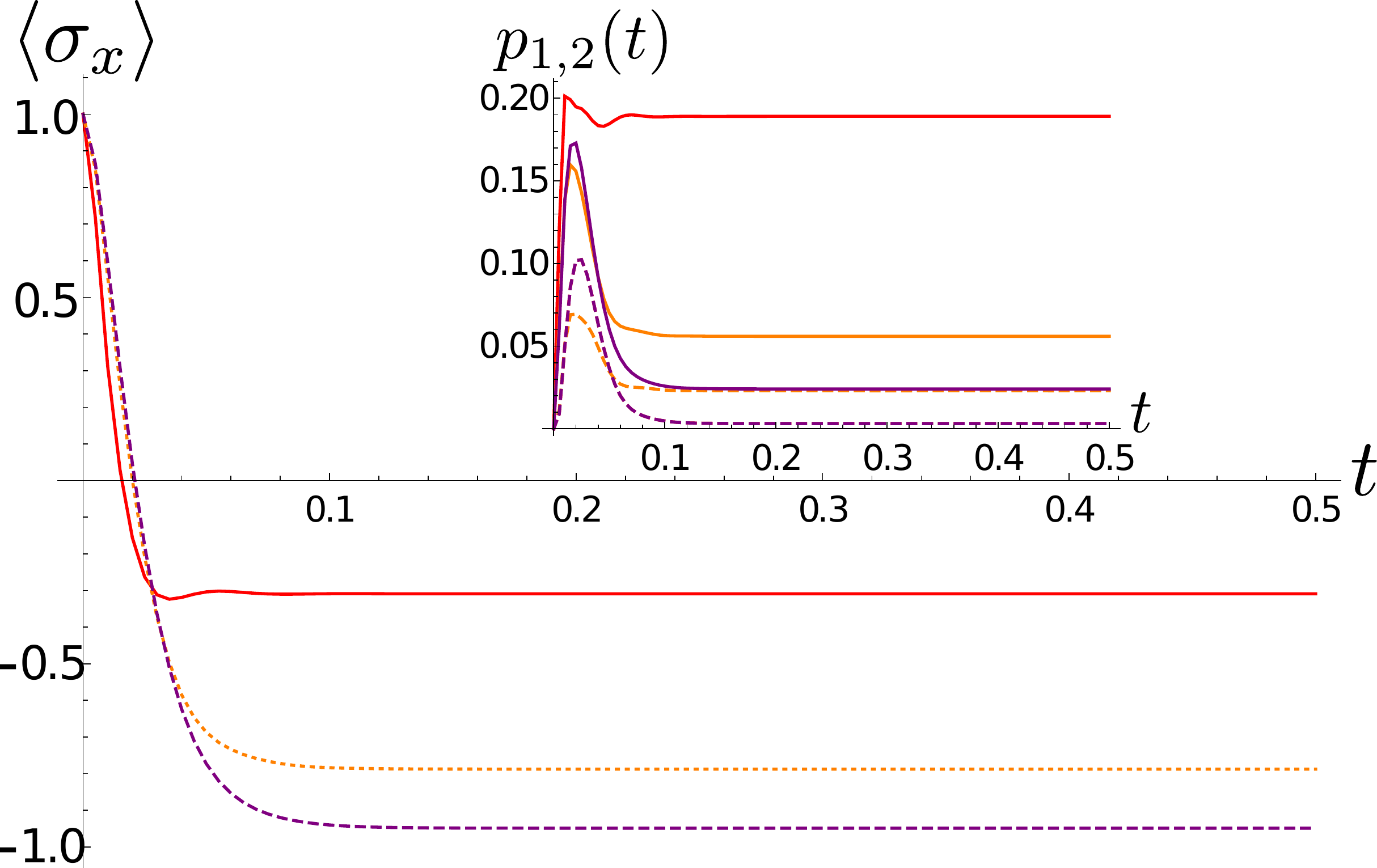}}
\subfigure[]{\includegraphics[width=0.25 \textwidth]{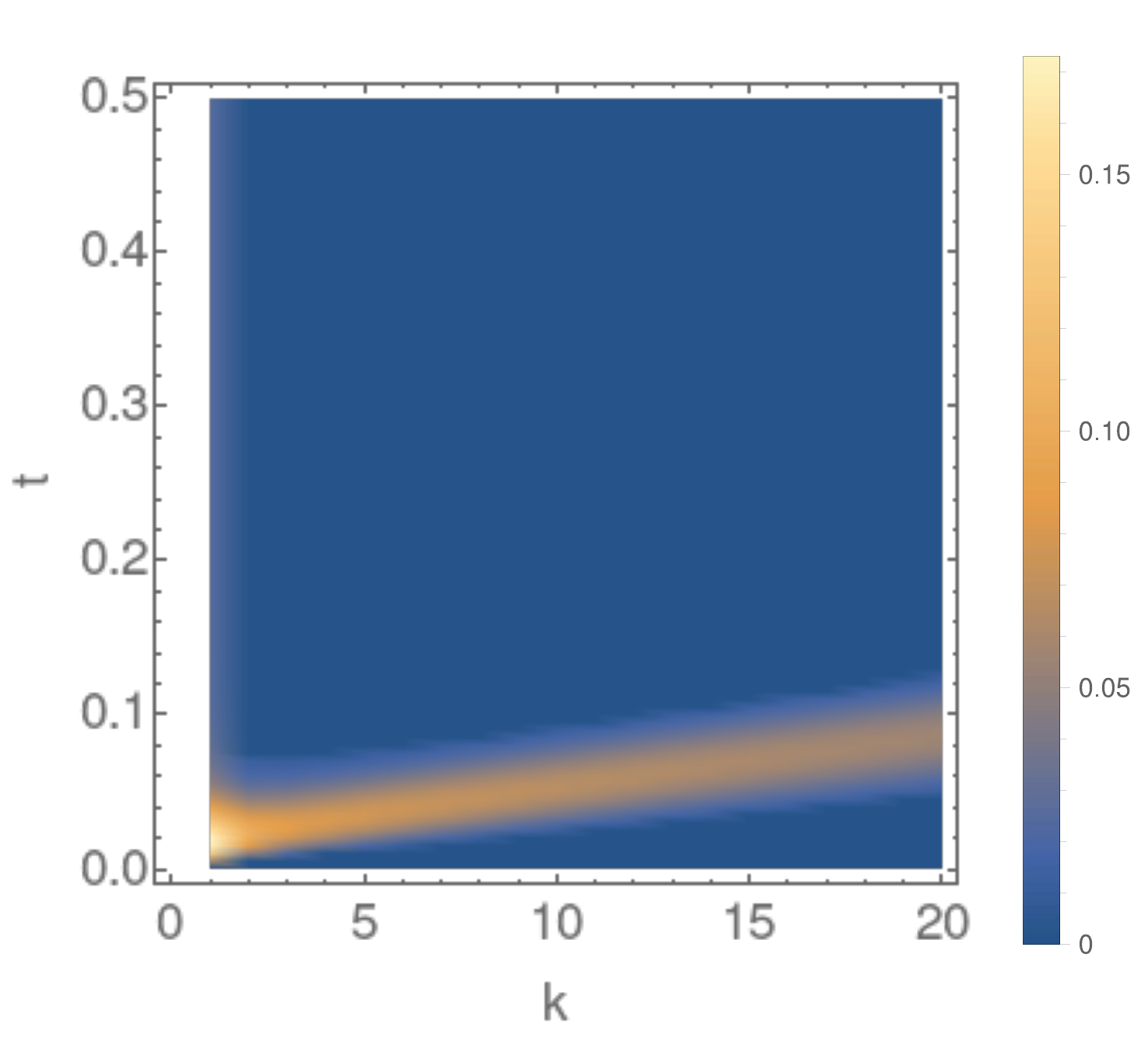}}
\subfigure[]{ \includegraphics[width=0.25 \textwidth]{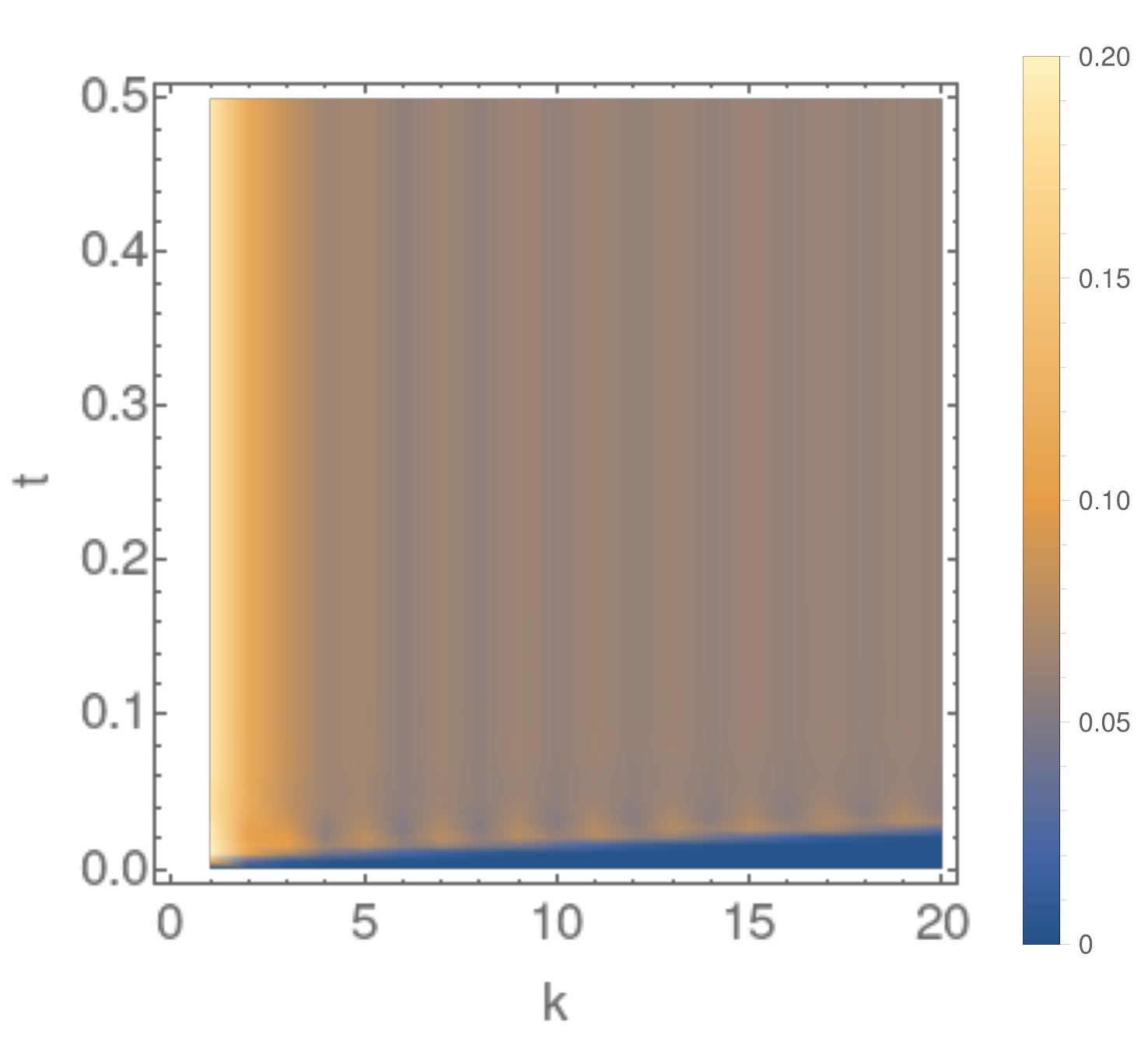}}

\caption{\label{fig:OhmicFulls2} Super-Ohmic SD ($s=2$).  Same quantities and line styles as in \Fref{fig:OhmicFulls05}.}
\end{figure}

Now we turn our attention to Ohmic spectral densities. As it happens for the Lorentian case discussed above, the main features of the excitations dynamics presented in Section \ref{sec:chaindyn} provides a key to understanding the results. We observed (see \Fref{fig:ohmicSDcoeff}(c) and (f)) that an excitation located at the first chain site will leave its initial location more slowly in the sub-Ohmic case than in the Ohmic and super-Ohmic case. Moreover, the excitation wavepacket tends for $s=0.5$ to be more spread over the chain than for $s=2$, with the case $s=1$ showing an intermediate behaviour. This features translate to the chain dynamics in the presence of an interaction with the system, as comparison between \Fref{fig:OhmicFulls05}, \ref{fig:OhmicFulls1} and \ref{fig:OhmicFulls2} shows. 

More in detail, we observe that at $T=0$ the excitations leave the first chain sites almost ballisticaly for $s=2$ (\Fref{fig:OhmicFulls2}(b)), whereas for $s=0.5$ there is an accumulation of excitations in the very first part of the chain (\Fref{fig:OhmicFulls05}(b)). The diagonal fringes appearing in the sub-Ohmic (and less pronounced in the Ohmic) case at zero temperature are easily explained in therms of the (moving in time) population profile shown in the inset of \Fref{fig:ohmicSDcoeff}(c). The inclination of the fringes, is instead related the the coupling coefficients beteween the TEDOPA chain oscillators that, as already pointed out, do not depend on $s$ but only on the spectral density support. 

At finite $T$ the situation changes quite drastically. First of all we observe that for all the chosen values of $s$ vertical fringes appear in frames (c) of \Fref{fig:OhmicFulls05}, \ref{fig:OhmicFulls1} and \ref{fig:OhmicFulls2}. Such vertical fringes can be associated to a the alternation of higher and lower average occupation number in nearest-neighbor sites, and allow to appreciate the onset of a stationary current when the state of the system gets close to its stationary state. A comparison between frames (a) of the same figures shows that in the sub-Ohmic the average occupation of the first TEDOPA chain sites is much higher than in the Ohmic and super-Ohmic cases. It must be noticed that, while the system-TEDOPA chain coupling $\kappa_0$ is equal for $T=0$ for all values of $s$, at finite temperature such coupling is inversely proportional to $s$. The sub-Ohmic TEDOPA chain is therefore more strongly coupled to the system, and this justifies the faster system dynamics at short times.


\section{Conclusion and outlook} \label{sec:Conclusion}
While chain mapping has been recognized as a powerful tool for the efficient simulation of open quantum system dynamics, the subtle role of excitation dynamics on the determination of such reduced dynamics has never been investigated in detail. This work represents a first step in this direction. While the single excitation dynamics is unable to capture the full complexity of the evolution of TEDOPA chains put in interaction with the system, it provides a most useful key to understand such evolutions, as in the case of the Lorentzian spectral density we considered. It moreover provides a mean to sensibly set  DMRG parameters, such as the chain truncation length and the local dimension of the chain oscillators: for super-Ohmic SDs, for example, the local dimension of the first TEDOPA chain oscillators must be set large enough as to host all the excitations that will accumulate in proximity of the system because of localization, while in the super-Ohmic case the local dimension of the first chain oscillators can be kept much smaller, since there is no signature of localization. While an analysis along the same lines for a specific spectral density was already presented \cite{tama19}, in this work we systematically compared and contrasted the features of the chain and full dynamics for a larger and very representative class of spectral densities. This allowed for example to shed light on the mechanisms allowing oscillators chain obtained by the T-TEDOPA procedure, and therefore starting from the vacuum state, to mimic an environment in the thermal state. For the Lorentzian case study such mechanism emerged quite clearly, and provided an key for the interpretation of the chain dynamics for SDs belonging to the Ohmic family. 

We moreover observed that,  while the asymptotic values of the TEDOPA coefficients determine the maximum distance reachable within a given time by an excitation initially located  at the beginning of the TEDOPA chain, or light-cone, the features of a specific spectral density are typically determined by a very small number of coefficients. Indeed, as it happens in the $\gamma=0.001$ Lorentzian SD case, the propagation of excitations in the light-cone can be hindered by an ``effective'' decoupling of the first sites of the chain from the remaining one. The analysis of the Ohmic SD instances, on the other side, showed that  different ($s$-dependent) chain coefficients in the very fist part of the chain lead to quite different occupation probability profiles of the sites within the light-cone. 

One of the, so far unexploited, advantages of chain mapping is the possibility of acquiring information on the state of the environment, something not meaningful when effective dynamics of Lindblad or Bloch-Redfield type are employed. While the number of chain modes perturbed by the interaction with the system is, in general, increasing with time, at any finite time it is in line of principle possible to make measurements on the oscillators in the light-cone. This could allow to understand, for example, which environmental modes are more involved in the dynamics and properly select the environmental reaction coordinates \cite{nazir18}. Moreover, in the presence of a fast convergence of the chain coefficients toward the asymptotic values, one expects a very small number of such coordinates. This represents a possible line of future research.
 
 There are features of the TEDOPA chain evolution that remained quite obscure. For example, the fringes that appear in the Ohmic scenario at finite temperature are not present in the Lorentzian case. Considered that, as already observed, for $\gamma/\Omega \ll 1$ a Lorentzian environment  can be assimilated to a damped harmonic oscillator undergoing a Lindblad-type dynamics, an therefore incoherently dissipating into an memoryless environment, one could read the lack of fringes as a signature of incoherent dynamics. A further analysis is therefore needed to better qualify the coherence dynamics in structured environments, and will be the focus of future work. 

\section*{Acknowledgements}
The author acknowledges most useful discussions with Andrea Smirne during the development of this work and has been supported by UniMi through the "Sviluppo UniMi" project.

%
\end{document}